# Unravelling the influence of shell thickness in organic functionalized $Cu_2O$ nanoparticles on $C_{2+}$ products distribution in electrocatalytic $CO_2$ reduction

*Jiajun Hu,[a] Silvio Osella,[b] Josep Albero[a]\* and Hermenegildo García[a]\**

J. Hu, J. Albero and H. García.
Instituto Universitario de Tecnología Química (CSIC-UPV), Universitat Politècnica de València (UPV), Avda. De los Naranjos s/n, 46022, Valencia, Spain.
E-mail: joalsan6@itq.upv.es; hgarcia@qim.upv.es

S. Osella.
Chemical and Biological Systems Simulation Lab, Centre of New Technologies, University of Warsaw, Warsaw, 02-097 Poland

Abstract: Cu-based electrocatalysts exhibits enormous potential for electrochemical $CO_2$ conversion to added-value products. However, high selectivity, specially towards $C_{2+}$ products, remains a critical challenge for its implementation in commercial applications. Herein, we report the preparation of a series of electrocatalysts based on octadecyl amine (ODA) coated $Cu_2O$ nanoparticles. HRTEM images show ODA coatings with thickness from 1.2 to 4 nm. DFT calculations predict that at low surface coverage, ODA tends to lay on the $Cu_2O$ surface, leaving hydrophilic regions. Oppositely, at high surface coverage, the ODA molecules are densely packed, being detrimental for both mass and charge transfer. These changes in ODA molecular arrangement explain differences in product selectivity. *In situ* Raman spectroscopy has revealed that the optimum ODA thickness contributes to the stabilization of key intermediates in the formation of $C_{2+}$ products, especially ethanol. Electrochemical impedance spectroscopy and pulse voltammetry measurements confirm that the thicker ODA shells increase charge transfer resistance, while the lowest ODA content promotes faster intermediate desorption rates. At the optimum thickness, the intermediates desorption rates are the slowest, in agreement with the maximum concentration of intermediates observed by *in situ* Raman spectroscopy, thereby resulting in a Faradaic efficiency to ethanol and ethylene over 73 %.

## 1. Introduction

Electrochemical $CO_2$ reduction (e$CO_2$R) is considered an appealing approach for the conversion of $CO_2$ emissions into high-value chemical products and fuels, especially in the periods in which there is a renewable electricity production surplus. However, one of the main bottlenecks that has hindered so far the commercial application of this technology is the lack of affordable, efficient, selective and stable electrocatalysts.[1]

Since Hori's seminal paper,[2] electrodes based on Cu and derivative compounds have been found among the most active ones for $CO_2$ conversion into different hydrocarbons and alcohols.[3] However, these Cu electrodes still suffer from low product selectivity and/or stability,[4] particularly at high current densities, limiting its commercial viability. In the case of Cu electrodes, the lack of product selectivity has been frequently attributed to the occurrence of competing side reactions, such as the hydrogen evolution reaction (HER), and to the moderate binding energy to *CO intermediates.[5] The last parameter is proposed to result in the simultaneous formation of one-carbon (C1) products, such as CO, $CH_4$, or HCOOH, together with variable proportions of two (or more)-carbon ($C_{2+}$) products, such as $CH_2=CH_2$, $CH_3CH_2OH$, $CH_3CH_2CH_2OH$, among others. The formation of liquid $C_{2+}$ products is generally more desirable than C1 products owing to their higher market price, energy density and their physical state at ambient conditions. Therefore, when designing Cu electrocatalysts for e$CO_2$R, it is essential to enhance the selectivity to the formation of $C_{2+}$ products by decreasing the occurrence of HER and the evolution of C1 products.[6]

To minimize HER, the initial step of e$CO_2$R, which involves the one-electron and one-proton reduction of $CO_2$ to form adsorbed *COOH, must be favored. Theoretical calculations have shown that the ratio of dissolved $CO_2$ to protons ($[CO_2]/[H^+]$) at the electrocatalyst/electrolyte interface significantly influences this reaction step.[7] Hence, to enhance selectivity toward e$CO_2$R over HER, it is advantageous to employ electrocatalysts with hydrophobic surface, which contribute to maintain a high ($[CO_2]/[H^+]$) ratio. On the other hand, since $C_{2+}$ product formation rate is proportional to the concentration of reactive C1 intermediates, such as *CO, *CHO, and *$CH_2$, a high surface coverage of reactive C1 intermediates is highly beneficial for efficient $C_{2+}$ product formation.[8]

Great progress at improving the selectivity of Cu-based electrocatalysts for e$CO_2$R has been achieved using alloys,[9] tuning the particle size[10] and morphology,[11] by control of the surface facets,[12] and doping with heteroatoms,[13] among other reported procedures.[14] More recently, the performance of heterogeneous Cu electrocatalysts has been improved by surface implantation of molecular modifiers that allow stabilization of key reaction intermediates and control surface hydrophobicity.[15] For example, Zhang et al have recently reported the preparation of $Cu_2O$ modified by cucurbit[6]urils for HER suppression. In this way, an enhancement of the Faradaic efficiency (FE) towards carbon products (CO + HCOOH) from 40 %, in the absence of the cucurbit[6]uril molecules, to 94 % in their presence has been reported. The authors attributed this enhancement to the strong $CO_2$ adsorption capacity on the surface and to its higher hydrophobicity, which simultaneously traps $CO_2$ near the $Cu_2O$ sites and limits $H_2O$ interaction.[16] In other example, Lim et al reported the selective $C_{2+}$ product formation with a FE of 77 %, on histidine-functionalised $Cu_2O$ electrocatalyst, decreasing HER FE to 21 %. The authors attributed this $eCO_2R$ selectivity to $C_{2+}$ products to the interaction of surface intermediates with histidine moieties that favors an increased intermediate population on the catalyst surface.[17] Besides molecular modifiers, Chen, et al. reported the coating of Cu electrodes with acrylamide-based polymers containing amino groups. A FE of 90 % towards $C_{2+}$ products was achieved at – 0.97 V vs. RHE, with only a < 7 % of FE towards HER. This high FE towards $C_{2+}$ products was ascribed to the pendant amino groups in the polymer backbone, which are proposed to lead to a higher concentration of $CO_2$ or carbonate species on the electrode surface, affecting also to the surface pH as a side effect.[18]

In those precedents, no attention has been paid to the influence on how molecular modifiers assemble on the electrodes surface and on the packing effect on the mass and charge transfer kinetics at the electrocatalyst interface. Herein, we have prepared a set of electrocatalysts for $eCO_2R$ based on octadecyl amine-functionalized $Cu_2O$ nanoparticles (NPs). Higher amounts of octadecyl amine (ODA) on the surface increase the ODA shell thickness around the $Cu_2O$ NPs, from 1.2 to 4 nm in average, affecting significantly to the $eCO_2R$ selectivity, there being an optimal thickness. Our DFT calculations have revealed that the reason for this selectivity dependence on ODA shell thickness is the molecule packing and position, laying on the $Cu_2O$ surface with incomplete surface coverage at low concentrations, while arranging perpendicular to the surface and densely packed at high contents. The packing and arrangement of ODA on the NP surface have a strong influence on the selectivity towards $C_{2+}$ products vs. undesirable competing HER. At the optimum shell thickness (2 nm), a 73 % of FE towards $C_{2+}$ products ($CH_2=CH_2$ and $CH_3CH_2OH$) and low HER (FE of 8 %) were measured at -0.9 V vs. RHE. Lower coverage results in enhanced HER due to the inefficiency of the thinner hydrophobic shell, while at larger ODA content, the concentration of key intermediates at the interface is also negatively affected due to impeded charge transfer. In situ Raman spectroscopy experiments confirmed that the concentration of key intermediates for ethylene and ethanol production is maximized at the optimum ODA thickness. In agreement, electrochemical impedance spectroscopy shows lower charge transfer resistance and high concentration of charged species at the optimum thickness, while at larger thickness promotes electrical insulation of the $Cu_2O$ surface, increasing charge transfer resistance. Similarly, pulse voltammetry measurements revealed larger charge accumulation in the samples with the lowest ODA content. However, at the optimum thickness, the rate of intermediates desorption is the slowest, favouring the $C_{2+}$ products formation. Finally, the activity of the ODA-functionalized $Cu_2O$ electrocatalyst has been evaluated in a flow cell system. $C_{2+}$ products such as ethylene, ethanol, propanol and acetic acid have been detected, achieving a FE to $C_{2+}$ of ca. 73% at 300 mA/cm$^2$ current density without apparent deactivation at this high current density.

## 2. Results and Discussion

### 2.1. ODA-functionalized $Cu_2O$ nanoparticles synthesis and characterization

$Cu_2O$ NPs coated by different amounts of ODA were prepared as illustrated in **Scheme 1**, following the synthetic procedure described in detail in the experimental section. In brief, colloidal $Cu_2O$ NPs were *in situ* formed by chemical reduction of Cu (II) acetylacetonate in octadecene (ODE) at 205 ºC in the presence of increasing ODA amounts. ODA is an amphiphilic molecule with a polar amino group head and an apolar C18 alkyl chain. The amino group has a dual function, providing a strong anchoring group to the $Cu_2O$ nanoparticles surface, as well as being a reducing agent during the $Cu_2O$ nanoparticles formation. Upon anchoring on the $Cu_2O$ surface the C18 alkyl chain renders hydrophobic the surface of the $Cu_2O$ NPs.

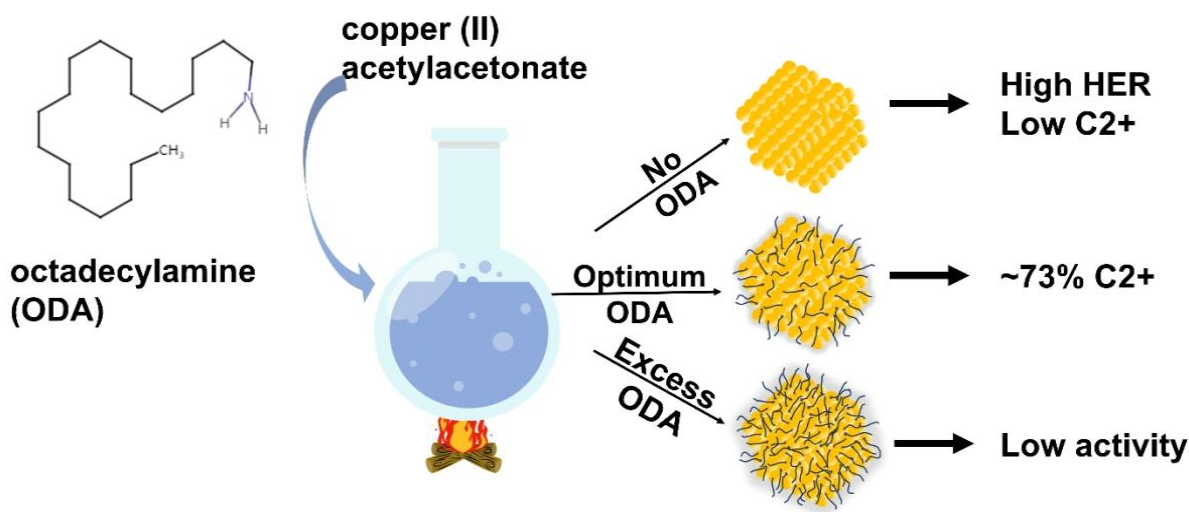

**Scheme 1.** Preparation procedure of ODA-functionalized Cu$_2$O nanoparticles.

The successful formation of Cu$_2$O NPs was confirmed by XRD and TEM. PXRD patterns presented in **Figure S1** in supporting information (SI) confirms Cu$_2$O as the only crystal phase present in all the samples. It is worth noticing that the diffraction peaks of the different samples broaden with the ODA content, which would be compatible with an amorphous shell growing in thickness with the increase in ODA loading during synthesis. The observed broadening is also evidenced by the slight increase in the lattice spacing values of the samples with ODA addition (**Table S1 in SI**). **Figure S2** in SI shows TEM images of the ODA-functionalized Cu$_2$O NPs at different ODA loadings. As can be observed in these images, all samples present similar polyhedric morphology and particle size distribution, with an average size of 130 nm, approximately.

The formation of an amorphous shell coating the Cu$_2$O NPs was confirmed by HRTEM (**Figure 1**). In these images, amorphous layers of different thicknesses can be observed in the samples under study. The average thickness of these shells has been calculated after measurement of a statistically relevant number of samples. In this way, thicknesses of $1.3 \pm 0.10$ nm, $2.0 \pm 0.12$ nm, $2.7 \pm 0.13$ nm and $4.0 \pm 0.20$ nm have been determined for CuODA1, CuODA2, CuODA3 and CuODA4, respectively.

Further confirmation of the ODA-functionalization on the Cu$_2$O nanoparticles surface has been obtained by ATR-FTIR spectroscopy (**Figure S3**). In all samples under study, the presence in the FTIR spectra of vibrational peaks located at 3328 cm$^{-1}$ attributable to N-H bond stretching was observed, together with peaks centred at 2914, 2845 and 719 cm$^{-1}$ that can be assigned to the asymmetric C-H bond stretch, the symmetric C-H stretching vibration, and the rocking mode, respectively. Moreover, the 1464 cm$^{-1}$ peaks have been attributed to C-H scissoring. Altogether, these FTIR spectra confirm the presence of ODA in all samples. In addition, vibration peaks at 623 cm$^{-1}$ in all samples can be assigned to Cu$_2$O, in good agreement with the PXRD data. It is worth noticing that the intensity of the ODA characteristic vibration peaks increases with the ODA loading. This is in good agreement with the thickness increase of the amorphous layers observed in the HRTEM images in Figure 1. Chemical composition analysis of the different samples was carried out by combining ICP-OES and combustion elemental analysis. The results are summarized in **Table S2** in SI. As can be seen there, addition of increasing amounts of ODA during the Cu$_2$O synthesis results in an enhancement of the C and N content in the final samples, being in accordance with the thickness increase of ODA on the Cu$_2$O nanoparticles surface.

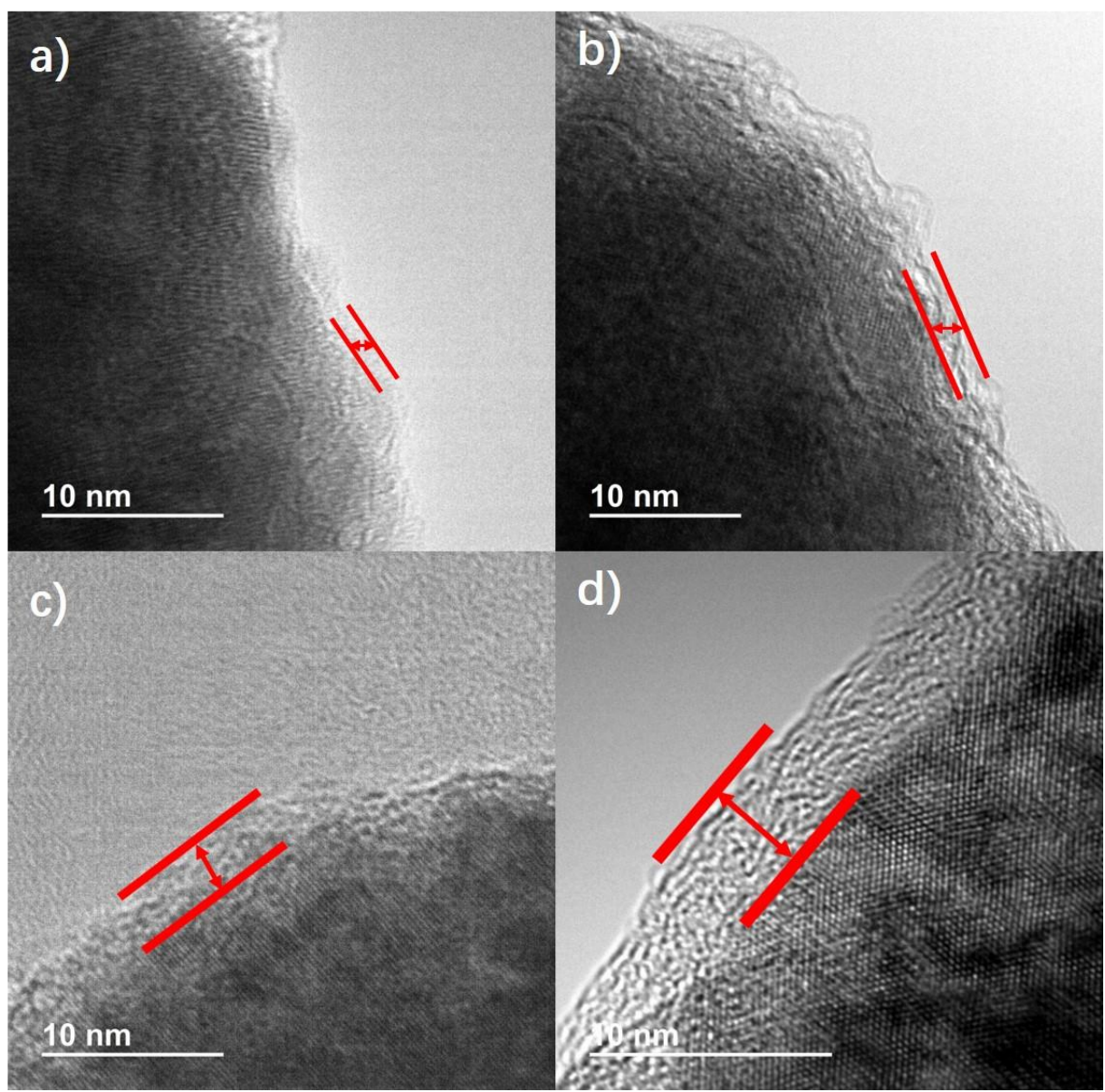

**Figure 1.** Representative HRTEM images of CuODA1(a), CuODA2(b), CuODA3(c) and CuODA4(d). The red arrows indicate the amorphous layer thickness.

The surface chemical composition of the as-prepared ODA-functionalized $Cu_2O$ NPs was further investigated by XPS. **Figure 2** shows the high resolution XPS data of Cu 2p 3/2 peaks and the best deconvolution to individual components for all samples. As can be observed, CuODA1 sample shows a single component centred at 932.68 eV, which can be ascribed either to $Cu^0$ or $Cu^I$ species. Alternatively, CuODA2, CuODA3 and CuODA4 spectra have been deconvoluted in two components. In these samples, a component located at 932.7 eV is also assigned either to $Cu^0$ or $Cu^I$. The second component, centred at 934.2 eV can be attributed to $Cu^{II}$ oxide species. It is worth noticing that $Cu^{II}$ crystal phases have not been detected by PXRD (see Figure S1 in SI) and, therefore, the $Cu^{II}$ species observed by XPS can only be attributed to a very thin, passivating layer on the $Cu_2O$ NPs surface. In addition, the relative contribution of the component assigned to $Cu^{II}$ species decrease with the ODA content, which is compatible with the role of ODA as reducing agent. In order to elucidate the actual Cu oxidation state of the first component, either $Cu^o$ or $Cu^I$, the Auger Cu LMM spectra was also acquired for all samples, the results being presented in **Figure S4** in SI.

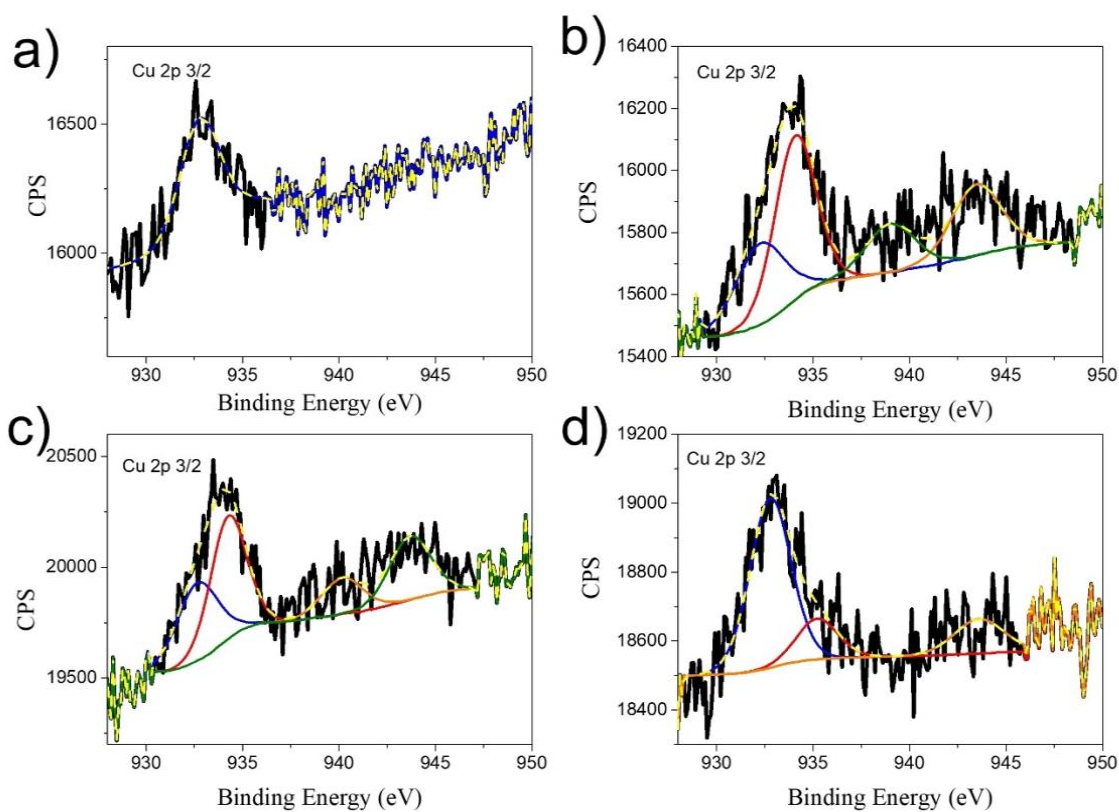

**Figure 2**. XPS core-level spectra of Cu 2p 3/2 of CuODA1 (a), CuODA2 (b), CuODA3 (c) and CuODA4 (d). Yelow lines correspond to the experimental data fitting, while the blue and red lines correspond to the $Cu^0$ and $Cu^I$ components. Orange and green lines correspond to the components related to satellites.

As can be seen there, while the Cu LMM Auger spectra of CuODA1 and CuODA4 show a single component related to $Cu^I$, the CuODA2 spectrum presents two components corresponding to $Cu^I$ and $Cu^{II}$, and the CuODA3 sample exhibits in addition a small third component attributed to $Cu^o$, besides $Cu^I$ and $Cu^{II}$. Overall, the as-prepared samples present different chemical environment at the NP surface depending on the ODA content. $Cu^I$ is common in all samples, in good agreement with XRD results. $Cu^{II}$ species are also present in most of the samples, especially in CuODA2 and CuODA3, while $Cu^0$ has been only observed in a small proportion in CuODA3. It should be noted that this surface composition belongs to the different samples prior the electrocatalytic reactions and, therefore, they could not correspond with the actual active catalyst form under the electrochemical reaction conditions, as reported in other precedent studies.[19]

## 2.2. Electrocatalytic $CO_2$ reduction reaction

The influence of ODA-coating on the $Cu_2O$ NPs surface in the $eCO_2RR$ was studied. For this purpose, glassy carbon electrodes supporting the different CuODA samples were prepared as described in detail in the experimental section and used as working electrodes in a H-type electrochemical cell in $CO_2$-saturated 0.1 M $KHCO_3$ electrolyte. Pt wire and KCl-saturated Ag/AgCl electrodes were used as counter and reference electrodes, respectively. A Nafion 117® membrane was used to separate the cathodic and anodic compartments.
All CuODA samples were active for $eCO_2RR$ (**Figure S5** in SI), but exhibiting different product distribution. Variable amounts of CO, $CH_4$, HCOOH, $CH_2CH_2$ and $CH_3CH_2OH$, together with $H_2$, were measured for the different ODA-functionalized $Cu_2O$ electrocatalysts. The partial product FE values for each of the samples were determined at different potentials in the range from -0.8 to -1.1 V vs. RHE. The results are presented in **Figure 3**. For all samples, the formation of carbon products is favoured at the lowest potentials (-0.8 and -0.9 V vs. RHE), while HER becomes predominant at higher potentials. The maximum selectivity towards $C_{2+}$ products ($CH_2=CH_2$ and $CH_3CH_2OH$) takes place at -0.9 V vs. RHE. Notably, at this potential, CuODA2 sample was the most selective electrocatalyst for $CH_2=CH_2$ and CH$_3$CH$_2$OH (FE of 17.8 ± 1.5 % and 55.5 ± 5.3 %, respectively), while FE of CO, CH$_4$, HCOOH and H$_2$ was only of 0.1 ± 0.0 %, 5.4 ± 0.5 %, 10.2 ± 1.2 % and 8.0 ± 0.7 %, respectively. The improved FE towards C$_{2+}$ products in CuODA2 is also evidenced by a higher partial current density towards C$_{2+}$, as can be observed in Figure 3 b. In order to put into a broad context the results herein obtained, a literature survey was carried out, and the FE values of reported precedents are summarized as **Table S3** in SI. As it can be there, CuODA2 performs among the best samples, only clearly surpassed by the polyamide incorporated Cu.

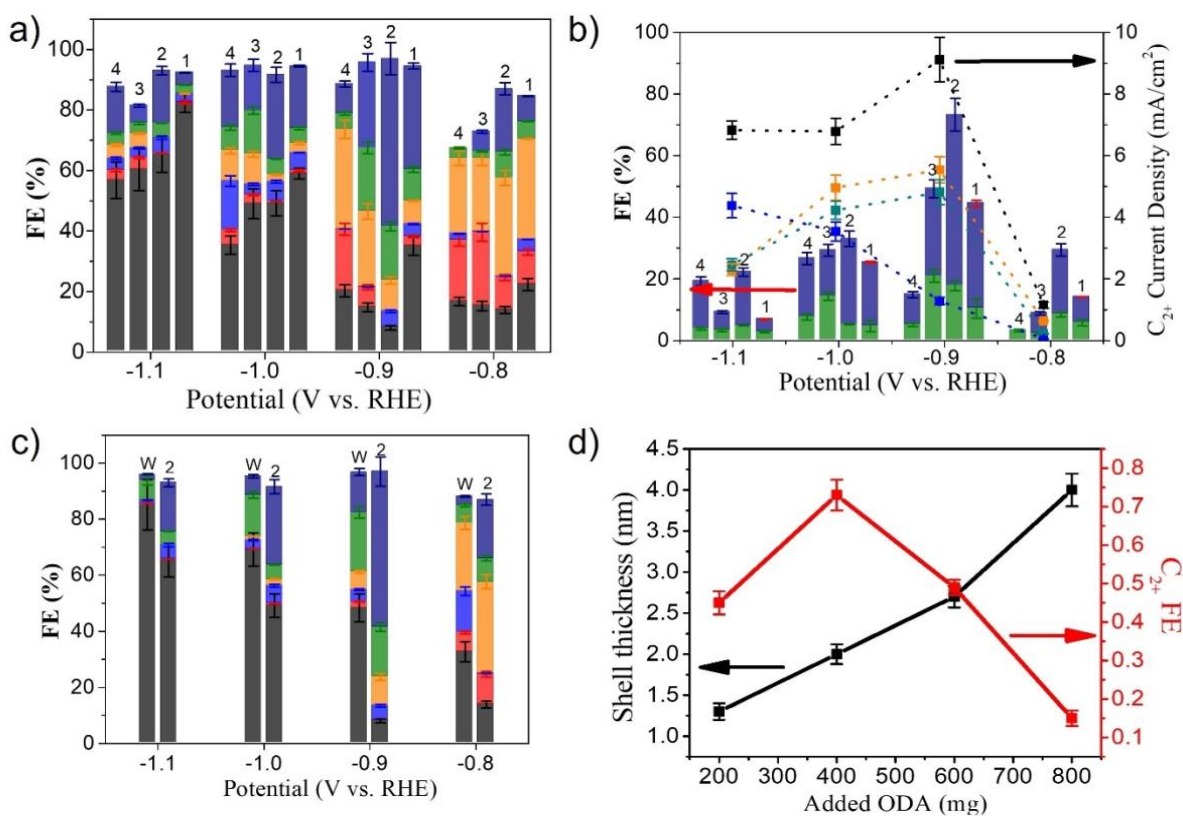

**Figure 3**. (a) FE for H$_2$ (grey), CO (red), CH$_4$ (blue), HCOOH (orange), CH$_2$=CH$_2$ (green) and CH$_3$CH$_2$OH (navy) obtained from CuODA1 (1), CuODA2 (2), CuODA3 (3) and CuODA4 (4) samples at different potentials. (b) FE (bars) and current density (squares) for C$_{2+}$ products. Green and navy bars correspond CH$_2$=CH$_2$ and CH$_3$CH$_2$OH, respectively. The C$_{2+}$ current density recorded from CuODA1, CuODA2, CuODA3 and CuODA4 samples is represented as orange, black, green and blue linked squares linked with dashed lines, respectively. (c) FE for H$_2$ (grey), CO (red), CH$_4$ (blue), HCOOH (orange), CH$_2$CH$_2$ (green) and CH$_3$CH$_2$OH (navy) evolution using CuODA2 (2) and the washed CuODA2 (W) samples at different potentials. (d) Comparison of the measured ODA thickness average (black line) and C$_{2+}$ FE (red line) with the added ODA. Error bars indicate standard deviation.

To understand the role of ODA coating, we submitted CuODA2 sample to a washing treatment in order to remove the ODA coating from the $Cu_2O$ surface (see experimental section) and, subsequently, the electrocatalytic activity of the washed sample (denoted as CuODA2-wash) was compared to that of the ODA-coated CuODA2. The successful removal of ODA from the $Cu_2O$ surface was confirmed by ATR-FTIR spectroscopy, ICP-OES and combustion elemental analysis (see **Figure S3** and **Table S1** in SI). As can be seen in ATR-FTIR, a drastic reduction of the characteristic ODA vibration peaks, together with a significant diminution in the C and N content were determined for CuODA2-wash. Meanwhile, the XRD pattern presented in **Figure S6** in SI shows that sample crystallinity has not been affected by the washing treatment, being $Cu_2O$ the only crystal phase detected. The polyhedric morphology and particle size is also preserved, as evidenced from the TEM images presented as **Figure S7** in SI. Moreover, we have studied the hydrophilic-hydrophobic properties of ODA coated $Cu_2O$ nanoparticles. **Figure S8** shows some photographs to illustrate the distinctive behaviour of CuODA2 and CuODA2-wash in water. Thus, CuODA2 exhibits a remarkable hydrophobicity, which can be visually observed by CuODA2 powder floating in water for 2 h. In contrast, CuODA2-wash underwent instantaneous precipitation under the same conditions.

Evaluation of the electrocatalytic activity of CuODA2-wash revealed that HER efficiency increases respect to CuODA2 at all the potentials under study (Figure 3 c). Accordingly, the $C_{2+}$ FE is significantly reduced at all potentials and, in particular, from 73.3 % to 35.5 at -0.9 V vs. RHE, while HER FE is enhanced from 8 to 48.3 %. This result clearly reveals the key role of ODA suppressing HER and favoring the selectivity towards $C_{2+}$ products.

Cyclic voltammetry of the samples at different scan rates (10 – 80 mV/s, **Figure S9** in SI) allowed us to determine the active electrochemical surface area (ECSA) by measuring the double-layer capacitance ($C_{DL}$). As can be observed in **Figure S10** in SI, the $C_{DL}$ decreased with the ODA content increase. Therefore, it seems that the Cu atoms at the NP surface become increasingly screened as the ODA thickness increases. Moreover, in agreement with this interpretation for the decrease of ECSA, the overall current density for the different CuODA samples also decreases with the ODA content, as depicted in the linear scanning voltammetry (LSV) plots in Figure S5 in SI. Thus, there is an optimum ODA thickness for the selective production of $C_{2+}$ products. Figure 3 d presents the relationship between the organic shell thickness measured from the HRTEM images and the $C_{2+}$ FE as a function of the added ODA during the sample preparation. The amorphous ODA layer thickness follows a nearly linear relationship with the ODA content present during the CuODA synthesis. In this way, an optimum thickness of ca. 2 nm for $C_{2+}$ products formation has been determined. An increase or diminution of the ODA layer thickness in just 0.5 nm, approximately, leads to a decrease in $C_{2+}$ FE and HER enhancement.

In order to determine the role of the ODA coating on the $CO_2$ adsorption capacity of these samples, the $CO_2$ adsorption isotherms of CuODA2 and CuODA2-wash were measured at 25 ºC, and the results are depicted as **Figure S11** in SI. As can be seen there, minor differences in $CO_2$ adsorption have been found, being the $CO_2$ adsorption capacity of CuODA2-wash slightly higher than that of CuODA2. This is indicating that the ODA coatings have not a significant role in the $CO_2$ adsorption, and the observed differences in FE towards C2+ products must be attributed to other factors as discussed below.

## 2.3. Investigation of the ODA thickness-selectivity relationship

In order to assess the influence of the ODA thickness on the product selectivity, density functional theory (DFT) calculations, in which different amounts of ODA were adsorbed on the $Cu_2O$ surface, were performed. In the models used for these calculations, a coverage of 17 % (corresponding to 3 ODA molecules adsorbed on our supercell system) leads to a thickness of 1.80 nm, while doubling the coverage to 33 % increases the thickness up to 2.14 nm, in the range of the optimal thickness observed by HRTEM for the selective $C_{2+}$ production in CuODA2. At lower ODA coverages, the ODA alkyl chains tend to maximize the interaction with the surface, lying flatter on it (see **Scheme 2**). However, low ODA coverage (< 17 %) still allows direct contact between the $Cu_2O$ surface and the aqueous electrolyte. This is experimentally evidenced by the higher ECSA obtained in CuODA1 and CuODA2-wash samples, containing the lowest ODA loadings (**Figure S10** in SI). As consequence, incomplete coverage would diminish the surface hydrophobicity of the sample, making easier the occurrence of HER, as depicted in **Figure 3** for CuODA1 and CuODA2-wash samples. On the opposite direction, at higher coverage percentage, the packing of the ODA molecules forces the alkyl chains to align almost perpendicularly to the surface, and very close to each other. However, a much too dense ODA shell affects negatively to the charge and mass transfer with the electrolyte. Hence, the FE towards $C_{2+}$ products obtained at the optimum ODA thickness of 2 nm would be a consequence of two opposite effects. An incomplete ODA coverage still allowing the occurrence of HER, while a too dense ODA shell being detrimental for the transfer of reagents or products with the surrounding medium as well as charge transfer with substrates.

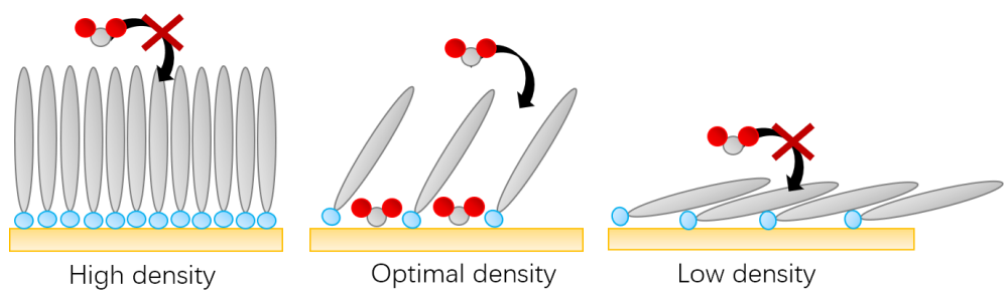

**Scheme 2.** Correlation between the ODA coverage and ability of $CO_2$ adsorption on the $Cu_2O$ substrate. At high density, the ODA are vertically aligned and tightly packed, hindering $CO_2$ adsorption; at low density the ODA molecules are almost flat on the surface, decreasing the ability of $CO_2$ to adsorb on the metal. At an optimal coverage, the ODA molecules are tilted with respect to the surface normal, allowing for $CO_2$ insertion.

Considering the 33 % coverage as optimal, data from the model being in accordance with the CuRN2 experimental system, we have performed a series of calculations on the adsorption of key intermediates which can lead to both ethylene and ethanol with this coverage percentage. Adsorption on the $Cu_2O$ surface at 33 % ODA coverage model was compared with calculations on a bare $Cu_2O$ surface (**Figure 4a**). In agreement with recent literature,[20] we considered the system charged. In particular, we focus on the *CHO and *CH adsorption as possible intermediates for $C_{2+}$ products, as well as one of the key intermediate to distinguish between the formation of ethylene (*CCH) or ethanol (*CCHOH). In fact, from the protonation of the carbon atoms in the *CCHOH intermediate, ethanol can be obtained in four steps, while protonation of *CCH should, eventually, lead to ethylene as product.
Our computational results show that the presence of ODA has a profound effect on the adsorption energy of the investigated intermediates, as it stabilizes the adsorption of $*CO_2$, *CCHOH and *CCH, while it destabilizes the *CHO and *CH adsorption in different degrees. The largest change due to the ODA influence is observed for the *CH adsorption energy, with a destabilization of 2.26 eV going from bare to ODA covered surface, due to the formation of a shorter bond in the first, of 0.182 nm, compared to a bond distance of 0.190 m, for the latter.
A different impact is observed for the *CCHOH and *CCH adsorption energies, being more stable when ODA is present with values of -0.81 and -0.78 eV, respectively, while on the bare surface the adsorption energy is weaker for both systems, with values of -0.46 and -0.62 eV, respectively. The weaker stability obtained for *CCHOH on the bare surface can be attributed to the lack of stabilizing interactions of the OH group in the intermediate with the nitrogen atoms present in the ODA molecules. This, in turn, has a profound impact for the production of either ethylene or ethanol. In fact, we can easily rationalize the experimental FE considering the reaction energies of the selected intermediates (Figure 4b). When ODA is present, the *CHO formation is thermodynamically favourable by -2.32 eV. From this intermediate, two pathways can be followed. The first leads to a double protonation of the carbonyl oxygen to obtain *CH, which can lead to ethylene after *CO insertion. Yet, *CH is thermodynamically unstable compared to *CHO, with an energy of -0.22 eV, thus hampering its formation and the subsequent ethylene formation. The second pathway considers the *CO insertion on the *CHO intermediate, leading to *CCHOH, with an energy of -2.03 eV, which is 0.1 eV more stable than the *CCH intermediate. This results in a slightly more favourable ethanol formation when ODA is present, through *CCHOH, rather than water elimination to obtain *CCH and eventually ethylene.

A different scenario arises when the bare surface is considered. In this case, *CH has been found more favourable thermodynamically compared to when ODA is present, although its energy is still higher than that of *CHO, with value of -2.27 eV. Yet, at the applied potential of -0.9 V vs. RHE there should be enough energy to overcome this barrier, and reaction can continue towards the ethylene formation. On the other hand, on bare $Cu_2O$ surface *CCH is more stable than *CCHOH by 0.1 eV, in a reverse situation than for the ODA covered surface, thus suggesting that now the ethylene path should be more favourable, in good agreement with the observed experimental data (Figure 3 c).

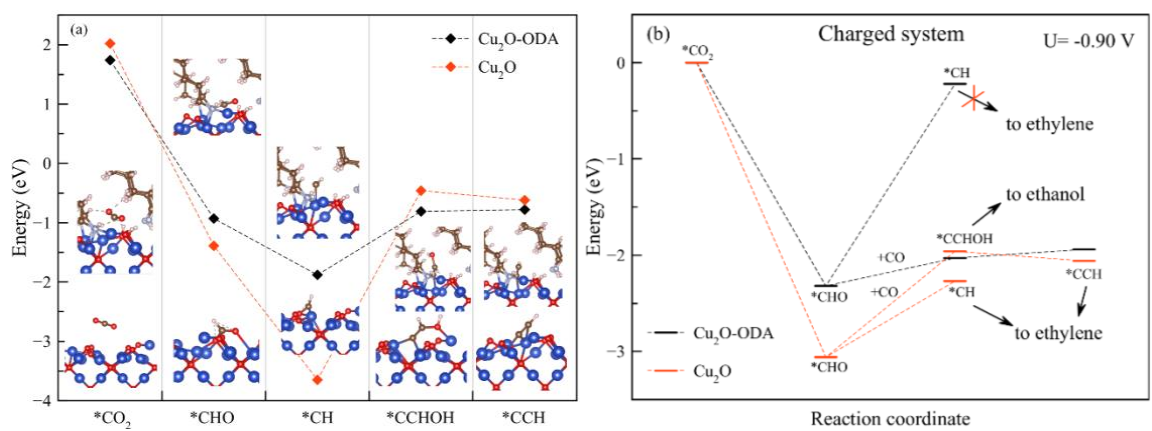

**Figure 4**. (a) Adsorption energy values of selected intermediates on the ODA covered (2.14 nm thickness) (black) and bare $Cu_2O$ surface (red). (b) reaction energy of selected intermediates at an applied potential of -0.9 V.

*In situ* electrochemical Raman spectroscopy studies were carried out for CuODA2, CuODA4 and CuODA2-wash samples, at open circuit potential (Voc) and at a cathodic potential of -0.9 V. The results are presented as **Figure S12** in SI. As it can be seen there, at -0.9 V potential the CuODA2 spectrum exhibits several peaks that were not detected in CuODA4 and CuODA2-wash. The main peak is located at 1537 $cm^{-1}$ that can be assigned to adsorbed carboxylate $*CO_2^-$ ,[21] which has been predicted to be one of the intermediates stabilized by ODA. Additional peaks at 1368 and 1676 $cm^{-1}$ have been previously attributed to carboxyl (*COOH) and bidentate $COO^-$ species and carboxylate group, respectively.[21b] In a high wavenumber region, two peaks centred at 2032 and 2089 $cm^{-1}$ can be observed both in CuODA2 and CuODA4, though in lesser intensity in the later. These have been previously attributed to *CO bonded at different sites.[21b, 22] In the proposed mechanism, this adsorbed *CO should insert in *CHO, leading to *CCHOH, which is the key intermediate in ethanol formation. The high intensity in the signals attributed to the *CO in CuODA2 than in CuODA4 suggests larger availability of adsorbed *CO that is required for the $C_{2+}$ products formation. The *CO intermediate could not be observed in the CuODA2-wash sample under these conditions.

Overall, DFT calculations and *in situ* electrochemical Raman spectroscopy support that the ODA shell at the optimum thickness contribute to the stabilization of key intermediates required for the formation of $C_{2+}$ products, especially ethanol. A decrease in the ODA content disminishes the ODA coverage on the $Cu_2O$ NPs, decreasing the hydrophobicity of the NP surface, and, consequently, favouring HER. On the contrary, a thicker ODA shell could prevent from efficient intermediates stabilization near the electrocatalyst surface, as well as forming an electrical insulating layer disfavouring charge and mass transfer from electrolyte to the Cu active sites.

To give additional support to our proposal, electrochemical impedance spectroscopy (EIS) was measured in all samples to disclose differences in the charge transfer properties of the samples. EIS spectra of the different samples were collected in $CO_2$-saturated 1 M $KHCO_3$ electrolyte across 20 different potentials between -0.2 and -0.8 V vs. RHE in 0.03 V increments. The results are presented as **Figure S13** in SI. As a representative example, the Nyquist plot of CuODA2 acquired at -0.7 V vs. RHE derived from these measurements is presented as **Figure S14** in SI.

As can be seen there, this plot is composed of a semicircle at higher frequency together with a second feature at lower frequencies, in good agreement with previously reported related examples.[17] The experimental data were fitted to a modified Randle's circuit model (see Figure S14 b), indicating two possible electrochemical interfaces with different time scales. The obtained charge transfer resistance at the two different time scales ($R_{LF}$ and $R_{HF}$ for the low frequency and high frequency domains, respectively) have been compared (**Figure 5**). $R_{HF}$ has been attributed to the charge transfer resistance at the electrolyte-electrocatalyst interface, while $R_{LF}$ is typically assigned to Faradaic reactions that take normally place in slower time scale. As can be seen, the main differences in $R_{LH}$ and $R_{HF}$ take place at low potentials. However, at -0.8 V vs. RHE, both $R_{HF}$ and $R_{LF}$ of all samples tend to converge into very similar values. This suggests favourable charge transfer during the reaction, in good agreement with previous reports.[17] CuODA4 and CuODA3 showed the highest $R_{LF}$, while and CuODA2 and CuODA2-wash demonstrated the lowest $R_{LF}$ of these measurements. This is not surprising due to the insulating behaviour of ODA, in good agreement with the $C_{DL}$ data and the DFT simulations. This is further evidenced by the low FE (60 -80 %) presented for CuODA3 and CuODA4 in **Figure 3**. Due to the insulating effect of the ODA coating in these samples, some charge could be consumed removing adsorbed species from the catalysts surface. Hence, some trace products have not not taken into account in the FE calculation.

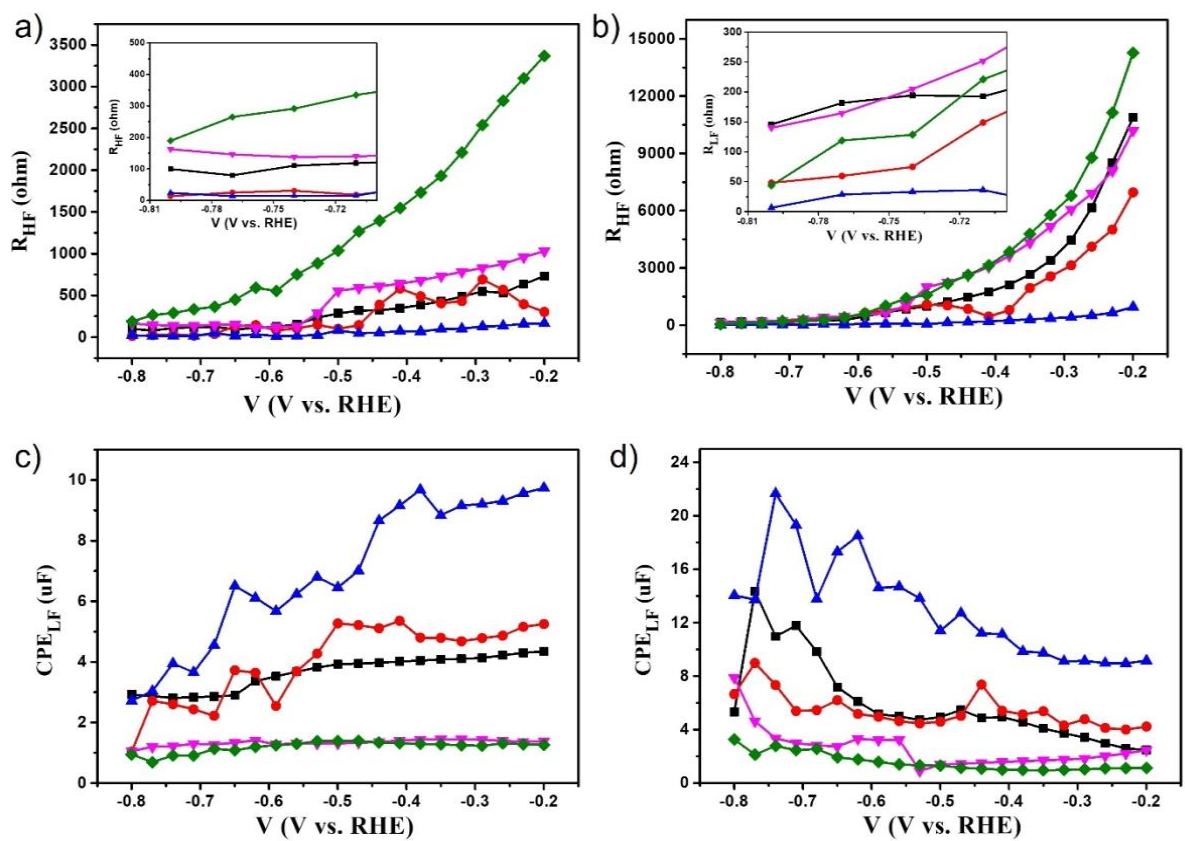

**Figure 5**. $R_{HF}$, $R_{LF}$, $CPE_{HF}$ and $CPE_{LF}$ values obtained from the best fitting of the Nyquist plots for CuODA1 (black squares), CuODA2 (red circles), CuODA3 (pink inverted triangles), CuODA4 (green diamond) and CuODA2-wash (blue triangle) in the potential range from – 0.2 to – 0.8 V vs. RHE.

On the other hand, the capacitive parameters, represented by the constant-phase elements $CPE_{HF}$ and $CPE_{LF}$, showed opposite trends. In the case of $CPE_{LF}$, the capacitance in all samples increased with the negative potential, indicating higher population of charged species in close contact with the catalysts surface. On the contrary, the $CPE_{HF}$ diminishes with the increase in cathodic potential, suggesting a lower population of charged species in the double layer. As can be seen, CuODA2-wash, CuODA1 and CuODA2, with lower ODA content, presented highest density of charged species on the catalyst surface, while CuODA3 and CuODA4, with the higher ODA thickness, exhibited the lowest population of charged species near the active sites. All-in-all pointing to the thicker ODA coating results in an increase of charge transfer resistance, and a concomitant decrease on the density of charged species on the electrocatalyst interface.

It is worth noticing, however, that the EIS measurements have been carried out between -0.2 and -0.8 V vs. NHE. Measurements with larger potentials, including the optimum -0.9 V vs.

RHE, are not possible due to large fluctuations in the capacitance values as consequence of the vigorous evolution of gases, which strongly disturb the EIS measurements. To overcome this experimental hurdle, we used pulse voltammetry that allowed us to analyse the charge accumulation and desorption kinetics on the electrocatalyst surface using short pulses (0.5 s) from an anodic potential of + 0.3 V vs NHE (near OCP) up to a cathodic potential range from -0.7 to -1 V vs. NHE (**Figure S15** a in SI), as reported before.[13] The transient current profiles of all samples were acquired in this way (see Figure S11 b-f in SI), and fitted to a combination of second-order and first-order decay kinetic functions (**Equation 1**).

$$F(t) = \frac{k_{2nd}}{(C_{2nd} + k_{2nd})^2} + C_{1st} \cdot e^{(-k_{1st} \cdot t)} \qquad \textbf{Equation 1}$$

From the experimental data fitting, second-order decay kinetics parameters ($C_{2nd}$ and $k_{2nd}$) were obtained for all samples. $C_{2nd}$ represents the initial current (t=0) and describes the surface charge accumulation, while $k_{2nd}$ describes the current decay rate and it is related with the cumulative desorption rate of species on the catalysts surface. As can be seen in **Figure 6**, $C_{2nd}$ followed identical trend than that obtained from $CPE_{HF}$ in Figure 5, where CuRN2-wash shows the highest initial surface accumulation, followed by CuODA1 and CuODA2, being CuODA3 and CuODA4 the samples with the lowest charge accumulation.

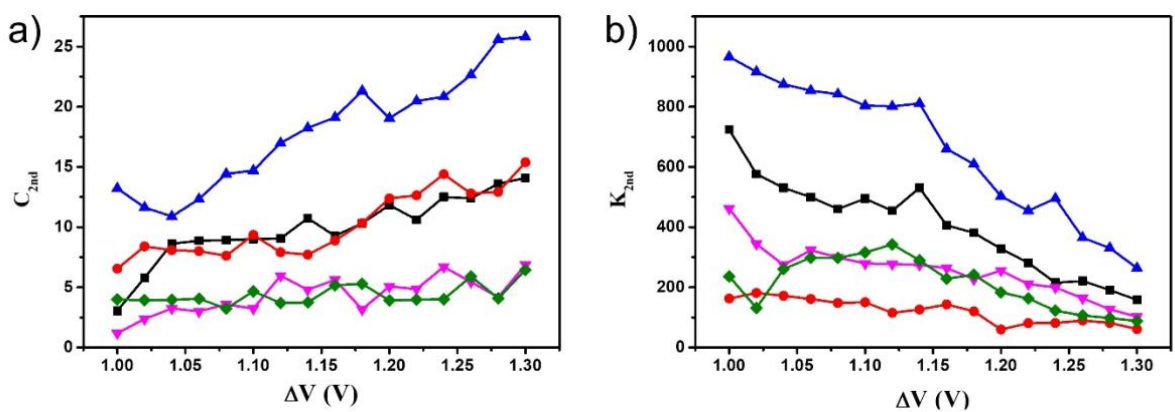

**Figure 6**. Pulsed voltammetry-derived parameters $C_{2nd}$ (a) and $k_{2nd}$ (b) obtained from the fitting of the pulse voltammetry measurements in CuRN1 (black squares), CuRN2 (red dots), CuRN3 (pink inverted triangles), CuRN4 (green diamonds) and CuRN2-wash (bleu triangles).

However, the $k_{2nd}$ parameter obtained from the different samples shows that CuODA2-wash exhibits the fastest desorption rate of species, while CuODA2 presents the slowest desorption kinetics. It is well known, that the formation $C_{2+}$ products require multiple electron and proton transfer, and therefore, intermediates should remain adsorbed on the surface catalysts long enough for this multiple charge and proton transfer. Hence, this result could explain the enhanced FE for $C_{2+}$ products demonstrated by CuODA2. In comparison with CuODA1, both present similar initial charge accumulation, however, the intermediate species desorb faster in CuODA1 than in CuODA2, explaining the higher selectivity towards $C_{2+}$ in CuODA2. CuODA3 and CuODA4 have lower initial surface charge than CuODA2, besides slightly faster desorption rates. Finally, although CuODA2-wash presents approximately double charge accumulation, the desorption rate is 5 times faster in CuODA2-wash than in CuODA2, and therefore, the intermediate species leading to $C_{2+}$ remain longer time on CuODA2 catalyst surface, promoting the improved FE for $C_{2+}$ products. Overall, pulsed voltammetry measurements allowed us determining that the slower intermediates desorption rate demonstrated in CuODA2 is the key factor resulting in a more efficient formation of $C_{2+}$.

**2.4. Evaluation of the electrocatalytic activity in flow electrochemical cell.**

It is well-known that the information obtained from the H-cell configuration cannot be extrapolated directly into the dynamic environment of a continuous flow electrolyser as consequence of the mass transport limitations in H-cell, which limit current densities to values lower than 100 mA/cm$^2$, bearing little resemblance with commercial electrolyzers. For these reasons, and encouraged by the outstanding selectivity towards $C_{2+}$ products obtained using CuODA2, the FE of this electrocatalyst was also tested under industrially relevant operation conditions, using a flow cell at the current density range from 100 to 400 mA/cm$^2$. The results are presented in **Figure S16** in SI. As can be seen there, FE towards $C_{2+}$ products increases with the current density up to 300 mA/cm$^2$, being the $C_{2+}$ FE of 73.5 %, very close to that obtained in the H-cell (Figure 3). At larger current density values (400 mA/cm$^2$), the $C_{2+}$ FE decreased to 60.3 %. Interestingly, $CH_3COOH$ and $CH_3CH_2CH_2OH$, beside $CH_2=CH_2$ and $CH_3CH_2OH$, were detected in the continuous flow conditions. For comparison purposes, CuODA2-wash sample was also evaluated under identical flow conditions. In this case, $CH_3COOH$, $CH_3CH_2CH_2OH$, $CH_2CH_2$ and $CH_3CH_2OH$ were also detected, but the $C_{2+}$ FE was of 42.6 %, in good agreement with the worse performance measured in the H-cell (Figure 3 c).

Finally, XRD pattern of CuODA2 after continuous flow operation at 300 mA/cm$^2$ for 2 h was compared to that of the fresh sample. The XRD presented in **Figure S17** in SI show that $Cu_2O$ is still the main crystal phase in the extensively used electrocatalyst, highlighting the stability of this electrocatalyst under industrially relevant conditions for the production of $C_{2+}$ products. Moreover, the stability of the ODA layer has been confirmed by HRTEM images, FITR spectroscopy and combustion elemental analysis of the used electrocatalyst. HRTEM images (**Figure S18** in SI) of CuODA2 after reaction show that the amorphous layer on the catalyst surface is preserved after reaction. Moreover, the FTIR spectrum after reaction (**Figure S19** in SI) presents identical vibrational peaks as the ones previously attributed to ODA. The combustion elemental analysis results are summarized in **Table S2** in SI. The obtained C and N contents (wt.%) in CuODA2 after the reaction are very similar to those of the fresh sample. However, a small decrease in C and N can be observed. This is suggesting that a small portion of the ODA molecules could have been released under the reaction conditions. Despite this, in overall, the CuODA2 electrocatalysts presents high stability under these reaction conditions.

## 3. Conclusion

The present results show that ODA coating is an efficient strategy to increase their FE of $Cu_2O$ NPs towards $C_{2+}$ products that can reach at the optimal shell thickness values of 73.6 % at 300 mA/cm$^2$. Electrochemical measurements indicates that this good performance derives from the balance between a slower desorption of the key reaction intermediates leading to $C_{2+}$ products and still adequate mass and charge transfer at the interface $Cu_2O$-electrolyte, while DFT calculations on models indicate that at the optimal loading ODA molecules tend to arrange perpendicularly to the $Cu_2O$ surface as consequence of the dense packing. ODA coating is stable at least for 2 h under high current density operation, due to the strong interaction of the Cu atoms with the amino groups. In this way, our results illustrate the high sensitivity of the product distribution in eCO2R to the surface of the electrocatalyst and the presence of simple molecules, showing a way for even further improvements.

## 4. Experimental Section

*Synthesis of CuODA electrocatalysts*: Firstly, a mixture of copper (II) acetylacetonate (0.130 g, 0.5 mmol), different amounts of octadecylamine (ODA) and 1-octadecene (ODE, 10 mL) was evacuated in a round bottom flask at room temperature and, then, heated to 135 ºC for 2 h. After complete dissolution of the reactants, the flask was filled with Ar, and the solution was heated to 205 ºC until the color changes to brown-yellow, allowing afterwards the reaction to continue for an additional 20 min. After this time, the system was cooled down at ambient temperature. Finally, the series of CuODAx samples were obtained after filtering, washing with toluene and drying at 60 °C overnight. CuODA1, CuODA2, CuODA3, and CuODA4 correspond to samples prepared using 200, 400, 600, and 800 mg of ODA, respectively. CuODA2-wash was obtained by sonicating the mixture solution of CuODA2 and $CHCl_3$ for 2 h.

*Characterization*: Powder X-ray diffraction (XRD) patterns were recorded on a Shimadzu XRD-7000 diffractometer by using Cu $\kappa_\alpha$ radiation (λ=1.5418 Å), operating at 40 kV and 40 mA at a scan rate of 10 ° per min in the 2–90° 2$\Theta$ range. The FTIR spectra were collected with a Bruker "Vertex 70" and a Thermo Nicolet 8700 spectrophotometer equipped with a DTGS detector (4 $cm^{-1}$ resolution, 32 scans). High-Resolution Transmission Electron Microscopy (HRTEM) images were obtained using a Philips CM300 FEG microscope operating at 200 kV, coupled with an X-Max 80 energy dispersive X-ray (EDX) detector (Oxford instruments). The microscope is equipped with the STEM unit, the dark-field and high-angle field (HAADF) image detectors. Inductively coupled plasma-optical emission spectrophotometry (ICP-OES, Varian 715-ES, CA, USA) analyses were carried out after completely dissolving the sample with nitric acid using calibrated standards. X-ray photoelectron spectra (XPS) were measured on a SPECS spectrometer equipped with a Phoibos 150 MCD-9 detector using a nonmonochromatic X-ray source (Al) operating at 200 W. Fitting of the experimental data to individual components was calculated from the area of the corresponding peaks after nonlinear background subtraction of the Shirley type. Atomic ratios of the different elements were determined from the areas of the corresponding XPS peaks, corrected by the response factor of the spectrometer. *In-situ* Raman spectra were obtained using a Renishaw "*in Via*" spectrophotometer equipped with an Olympus optical microscope. Electrochemical characterization was carried out in a Gamry instruments model 5000E.

*Electrocatalytic $CO_2$ reduction tests*: A circular glassy carbon electrode with 3 mm diameter was used as the working electrode. 5 mg of the catalyst was dispersed in 0.5 mL isopropanol and 40 μL Nafion solution via ultrasonication in an ice bath for 40 min. 30 μL of the catalyst ink was deposited onto the polished glassy carbon to achieve a 0.42 mg/$cm^2$ catalyst loading. The catalyst ink was then dried overnight at room temperature. All electrochemical measurements were done using a calibrated potentiostat (Gamry Interface 5000E) using an Ag/AgCl electrode with saturated KCl as the reference, platinum wire and 0.1 M $KHCO_3$ as the counter electrode and electrolyte, respectively. RHE potentials were calculated based on the Nernst equation ($E_{RHE}$ = $E_{Ag/AgCl}$ + 0.0592 × pH + 0.197 V). The products and Faradaic efficiency of $CO_2$ reduction were measured using chronoamperometry at a fixed potential in a H-type electrochemical cell separated by a Nafion membrane. An open cell without membrane was used for pulsed voltammetry and EIS measurements. Pulsed voltammetry was conducted in 1 M $KHCO_3$. The anodic potential was set around 0.3 V, where Faradaic processes were minimal. The cathodic potential is varied at 0.02 V intervals, with ΔV values (ΔV = $V_{anodic}$ − $V_{cathodic}$) ranging from 1 V to 1.3 V. Both anodic and cathodic pulses were applied for 20 s and the sampling time of the current was 0.003 s. The anodic current decays were fitted from t = 0 s to t = 3 s with positive bounds.

Electrochemical impedance spectroscopy (EIS) data were collected with 1 M $KHCO_3$, with higher electrolyte concentration used to minimize resistance as well as create a more stable EIS environment. The spectra were then collected across 20 different potentials between the range of -0.2 to -0.8 V vs. RHE in 0.03 V increments. The spectra were collected from 0.5 Hz to 30 kHz at 10 points per decade. The obtained Nyquist plots were individually fitted using the Simplex method on Gamry Echem Analyst.

The evolved gases were analyzed using a gas chromatograph (Agilent 7890A) equipped with Carboxene 1010 column analysing $CO_2$, CO and up to $C_4$ hydrocarbons, and a thermal conductivity detector (TCD) for $H_2$. Quantification of the percentage of each gas was based on prior calibration of the system injecting mixtures with known percentage of gases. The liquid products were detected using Bruker AVANCE III 400 MHz nuclear magnetic resonance (NMR). The Faradic efficiency of $H_2$ and CO generation was calculated according to the following equation:

FE=n*z*F/Q

where n denotes the moles of the generated products, z represents the number of electrons exchanged for products formation, F = 96485 C·mol$^{-1}$ represents the Faraday constant, and Q is the total charge.

For flow cell measurements, 1M KOH aqueous solution was used as electrolyte and Pt/Ti plate and catalyst-coated gas diffusion layers (5 cm$^2$) were used as counter and working electrodes, respectively. The anodic and cathodic chambers were separated by an anion exchange membrane (Fumatech FAAM-PK-75). The flow rate of $CO_2$ was 25 mL/min, and the catholyte was circulated at 20 mL/min using a peristaltic pump (REDOX.ME). The performance of $CO_2$ reduction at different currents was evaluated by chronopotentiometry.

*DFT calculations*: All geometry optimizations calculations were performed using spin-polarized density functional theory (DFT) as implemented in the Vienna ab initio simulation package (VASP).[23] The Perdew-Burke-Emzerhof (PBE) functional with a plane-wave cutoff energy of 500 eV was used. A 3x2x2 $Cu_2O$ supercell was constructed in order to accommodate 6 ODA molecules on it (translating into a coverage of 33%), which reflects the thickness of the ODA layer experimentally observed. Due to the large size of the supercell used, the Brillouin zone was sampled using a $1 \times 1 \times 1$ k-point grid centered at Gamma. A vacuum space of 1 nm in the *z* direction (perpendicular to the basal plane) was used to avoid interactions between periodic images. The convergence criteria for the force on each atom was set to 0.05 eV/Å, while the electronic structure energy convergence criteria was $10^{-5}$ eV. The Grimme D3 method with Becke-Johnson parameters[24] were employed to account for Van der Waals interactions.[25] Different gas molecules such as water, $CO_2$ and CO as well as key reaction intermediates (CHO, CH, CCHOH, CCH) have been considered for adsorption on the $Cu_2O$-ODA and $Cu_2O$ bare surfaces. The adsorption energy has been calculated according to the following equation:

$$E_{ads} = E_{cpl} - E_{surf} - E_{int}$$

Where $E_{cpl}$ is the energy of the adsorbed species on the surface, $E_{surf}$ is the energy of the Cu2O surface (with or without ODA) and $E_{int}$ is the energy of the intermediate species in gas phase.

The reaction energy was computed based on the computational hydrogen electrode (CHE) model, using the following Equation:

$$\Delta E = \Delta E_{DFT} + kTln10 \; x \; pH - eU$$

where $\Delta E_{DFT}$ is the total energy from DFT simulations, pH=7 and U is the applied electrode potential (-0.9 V).

**Supporting Information**

Supporting Information is available from the Wiley Online Library or from the author.

**Acknowledgements**

Financial support by the Spanish Ministry of Science and Innovation (CEX-2021-001230-S and PDI2021-0126071-OB-CO21 funded by MCIN/AEI/ 10.13039/501100011033) and

Generalitat Valenciana (Prometeo 2021/038 and Advanced Materials programme Graphica MFA/2022/023 with funding from European Union NextGenerationEU PRTR-C17.I1). Participation in the EU project ECO2Fuel is grateful acknowledged. J.A. thanks the Spanish Ministry of Science and Innovation for a Ramon y Cajal research associate contract (RYC2021–031006-I financed support by MCIN/AEI/10.13039/501100011033 and by European Union/NextGenerationEU/ PRTR), and Generalitat Valenciana (CIGE 2022-093) financed by European Union-Next Generation EU, through the Conselleria de Innovación, Universidades, Ciencia y Sociedad Digital. J.H. thanks the Chinese Scholarship Council for doctoral fellowship. S.O. acknowledges the National Science Centre, Poland (grant no. UMO/2020/39/I/ST4/01446) and the "Excellence Initiative – Research University" (IDUB) Program, Action I.3.3 – "Establishment of the Institute for Advanced Studies (IAS)" for funding (grant no. UW/IDUB/2020/25). The computation was carried out with the support of the Interdisciplinary Center for Mathematical and Computational Modeling at the University of Warsaw (ICM UW) under grants no. G83-28 and GB80-24.

Received: ((will be filled in by the editorial staff))
Revised: ((will be filled in by the editorial staff))
Published online: ((will be filled in by the editorial staff))

References

[1] M. B. Ross, P. De Luna, Y. Li, C.-T. Dinh, D. Kim, P. Yang, E. H. Sargent, *Nature Catalysis* **2019**, 2, 648.
[2] Y. Hori, A. Murata, R. Takahashi, *Journal of the Chemical Society, Faraday Transactions 1: Physical Chemistry in Condensed Phases* **1989**, 85, 2309.
[3] M. K. Birhanu, M.-C. Tsai, A. W. Kahsay, C.-T. Chen, T. S. Zeleke, K. B. Ibrahim, C.-J. Huang, W.-N. Su, B.-J. Hwang, **2018**, 5, 1800919.
[4] F. Dattila, R. R. Seemakurthi, Y. Zhou, N. López, *Chemical Reviews* **2022**, 122, 11085.
[5] S. Nitopi, E. Bertheussen, S. B. Scott, X. Liu, A. K. Engstfeld, S. Horch, B. Seger, I. E. L. Stephens, K. Chan, C. Hahn, J. K. Nørskov, T. F. Jaramillo, I. Chorkendorff, *Chemical Reviews* **2019**, 119, 7610.
[6] T. Kim, G. T. R. Palmore, *Nature Communications* **2020**, 11, 3622.
[7] N. Gupta, M. Gattrell, B. MacDougall, *Journal of Applied Electrochemistry* **2006**, 36, 161.
[8] T. Cheng, H. Xiao, W. A. Goddard, *Journal of the American Chemical Society* **2017**, 139, 11642.
[9] D. Wei, Y. Wang, C.-L. Dong, Z. Zhang, X. Wang, Y.-C. Huang, Y. Shi, X. Zhao, J. Wang, R. Long, Y. Xiong, F. Dong, M. Li, S. Shen, **2023**, 62, e202217369.
[10] H. Jung, S. Y. Lee, C. W. Lee, M. K. Cho, D. H. Won, C. Kim, H.-S. Oh, B. K. Min, Y. J. Hwang, *Journal of the American Chemical Society* **2019**, 141, 4624.
[11] a) Y. Jiang, X. Wang, D. Duan, C. He, J. Ma, W. Zhang, H. Liu, R. Long, Z. Li, T. Kong, X. J. Loh, L. Song, E. Ye, Y. Xiong, **2022**, 9, 2105292; b) D. Wakerley, S. Lamaison, F. Ozanam, N. Menguy, D. Mercier, P. Marcus, M. Fontecave, V. Mougel, *Nature Materials* **2019**, 18, 1222.

[12] J. Li, Y. Kuang, Y. Meng, X. Tian, W.-H. Hung, X. Zhang, A. Li, M. Xu, W. Zhou, C.-S. Ku, C.-Y. Chiang, G. Zhu, J. Guo, X. Sun, H. Dai, *Journal of the American Chemical Society* **2020**, 142, 7276.
[13] G. O. Larrazábal, A. J. Martín, F. Krumeich, R. Hauert, J. Pérez-Ramírez, **2017**, 10, 1255.
[14] S. Wang, T. Kou, S. E. Baker, E. B. Duoss, Y. Li, *Materials Today Nano* **2020**, 12, 100096.
[15] D.-H. Nam, P. De Luna, A. Rosas-Hernández, A. Thevenon, F. Li, T. Agapie, J. C. Peters, O. Shekhah, M. Eddaoudi, E. H. Sargent, *Nature Materials* **2020**, 19, 266.
[16] Y. Zhang, X.-Y. Zhang, K. Chen, W.-Y. Sun, *ChemSusChem* **2021**, 14, 1847.
[17] C. Y. J. Lim, M. Yilmaz, J. M. Arce-Ramos, A. D. Handoko, W. J. Teh, Y. Zheng, Z. H. J. Khoo, M. Lin, M. Isaacs, T. L. D. Tam, Y. Bai, C. K. Ng, B. S. Yeo, G. Sankar, I. P. Parkin, K. Hippalgaonkar, M. B. Sullivan, J. Zhang, Y.-F. Lim, *Nature Communications* **2023**, 14, 335.
[18] X. Chen, J. Chen, N. M. Alghoraibi, D. A. Henckel, R. Zhang, U. O. Nwabara, K. E. Madsen, P. J. A. Kenis, S. C. Zimmerman, A. A. Gewirth, *Nature Catalysis* **2021**, 4, 20.
[19] T. Möller, F. Scholten, T. N. Thanh, I. Sinev, J. Timoshenko, X. Wang, Z. Jovanov, M. Gliech, B. Roldan Cuenya, A. S. Varela, P. Strasser, **2020**, 59, 17974.
[20] A. K. Buckley, M. Lee, T. Cheng, R. V. Kazantsev, D. M. Larson, W. A. Goddard Iii, F. D. Toste, F. M. Toma, *Journal of the American Chemical Society* **2019**, 141, 7355.
[21] a) I. V. Chernyshova, P. Somasundaran, S. Ponnurangam, *PNAS* **2018**, 115, E9261; b) S. Jiang, K. Klingan, C. Pasquini, H. Dau, *The Journal of Chemical Physics* **2018**, 150.
[22] a) D. Zhong, Z.-J. Zhao, Q. Zhao, D. Cheng, B. Liu, G. Zhang, W. Deng, H. Dong, L. Zhang, J. Li, J. Li, J. Gong, **2021**, 60, 4879; b) K. Yao, J. Li, H. Wang, R. Lu, X. Yang, M. Luo, N. Wang, Z. Wang, C. Liu, T. Jing, S. Chen, E. Cortés, S. A. Maier, S. Zhang, T. Li, Y. Yu, Y. Liu, X. Kang, H. Liang, *Journal of the American Chemical Society* **2022**, 144, 14005.
[23] a) P. E. Blöchl, *Physical Review B* **1994**, 50, 17953; b) G. Kresse, J. Furthmüller, *Physical Review B* **1996**, 54, 11169; c) G. Kresse, J. Furthmüller, *Computational Materials Science* **1996**, 6, 15.
[24] S. Grimme, S. Ehrlich, L. Goerigk, **2011**, 32, 1456.
[25] S. Grimme, J. Antony, S. Ehrlich, H. Krieg, *The Journal of Chemical Physics* **2010**, 132.

Cu$_2$O nanoparticles have been functionalized with octadecyl amine shells of different thicknesses, and their electrocatalytic activity for CO$_2$ reduction evaluated. At the optimum shell thickness, a 73 % Faradaic efficiency towards C2+ products (ethylene and ethanol) has been obtained. DFT calculations, *in situ* Raman spectroscopy, electrochemical impedance and pulse voltrametry have been employed to stablish structure-selectivity relationships.

Jiajun Hu, Silvio Osella, Josep Albero* and Hermenegildo García*

**Unravelling the influence of shell thickness in organic functionalized Cu$_2$O nanoparticles on C2+ products distribution in electrocatalytic CO$_2$ reduction**

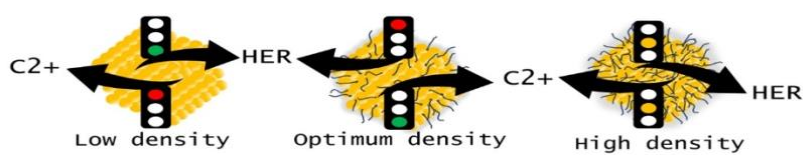

**Unravelling the influence of shell thickness in organic functionalized $Cu_2O$ nanoparticles on C2+ products distribution in electrocatalytic $CO_2$ reduction**

*Jiajun Hu, Silvio Osella, Josep Albero\* and Hermenegildo García\**

**Table S1**. Chemical composition of the samples under study, obtained from ICP-OES and elemental analysis.

| Angle (2 θ) | CuODA1 (Å) | CuODA2 (Å) | CuODA3 (Å) | CuODa4 (Å) |
|---|---|---|---|---|
| 29.7400 | 3.0041 | 3.0001 | 3.0072 | 3.0109 |
| 36.6279 | 2.4535 | 2.4586 | 2.4594 | 2.4614 |
| 42.4921 | 2.1275 | 2.1211 | 2.1253 | 2.1286 |
| 61.6506 | 1.5045 | 1.5039 | 1.5059 | 1.5054 |
| 73.7942 | 1.2841 | 1.2834 | 1.2840 | 1.2855 |

**Table S2**. Chemical composition of the samples under study, obtained from ICP-OES and elemental analysis.

|  | Cu (wt.%) | O (wt.%) | C (wt.%) | N (wt.%) | H (wt.%) |
|---|---|---|---|---|---|
| CuODA1 | 56.97 | 20.21 | 18.52 | 1.23 | 3.07 |
| CuODA2 | 48.86 | 19.37 | 25.85 | 1.51 | 4.41 |
| CuRNODA | 45.28 | 13.51 | 33.87 | 1.73 | 5.61 |
| CuRNODA4 | 32.28 | 8.46 | 48.26 | 2.73 | 8.32 |
| CuODA2-Wash | 83.74 | 13.53 | 2.18 | 0.25 | 0.3 |
| CuODa2 used | 54.66 | 22.80 | 21.05 | 1.15 | 1.34 |

**Table S3**. Comparison of the $eCO_2RR$ performance of reported cathodic electrocatalysts with the activity of CuODA2

| Entry | Catalyst | Electrolyte | Potential/Current | C1 FE (%) | C2+ FE(%) | $H_2$ FE(%) | Ref. |
|---|---|---|---|---|---|---|---|
| 1 | Cu dendrite | $CO_2$-saturated 0.1 M $CsHCO_3$ | From -1.1 V to -1.5 V vs. RHE (-30 mA) | 11 | 73 | 10 | [1] |
| 2 | Cucurbit[6]urils-modified $Cu_2O$ | 0.5 M $KHCO_3$ | -0.7 V vs. RHE | 94 | 0 | 6 | [2] |
| 3 | Polyamide-incorporated Cu | 1 M KOH | -0.97 V (433 mA/cm$^2$) | ~7 | 90 | ~3 | [3] |
| 4 | Histidine-functionalized Cu | 1 M $KHCO_3$ | -2 V vs. RHE | < 5 | 77 | 21 | [4] |
| 5 | Tetrahydro-bipyridine-functionalized Cu | 1 M $KHCO_3$ | -0.83 V vs. RHE | - | 72 | - | [5] |
| 6 | CuODA2 | 0.1 M $KHCO_3$ | -0.9 V vs. RHE | 15.7 | 73.3 | 11 | This work |
| 7 | N-substituted pyridinium Cu | 0.1 M $KHCO_3$ | -1.1 V vs. RHE | - | 70-80 % | - | [6] |
| 8 | Phenanthrolinium-derived Cu | 0.1 M $KHCO_3$ | -4.4 V full-cell potential | - | 66 | - | [7] |
| 9 | FeTPP[Cl] inmobilized Cu | 1 M $KHCO_3$ | -0.82 V vs. RHE | - | 41 | - | [8] |
| 10 | Dendritic polymer amine terminated-Cu | 1 M KOH | -0.97 V vs. RHE | 26 | 69 | < 5 | [9] |
| 11 | Bipyridine film on Cu | 1 M $KHCO_3$ | -0.96 V vs. RHE | 3 | 46 | 25 | [10] |
| 12 | Thiazole functionalized Ag-Cu | 0.5 M $KHCO_3$ | -4.5 V cell Voltage (250 mA/cm$^2$) | 10 | 80 | 5.3 | [11] |
| 13 | Tannic acid modified Cu | 1 M KOH | -1.2 V vs. RHE | 30 | 63.6 | 10 | [12] |

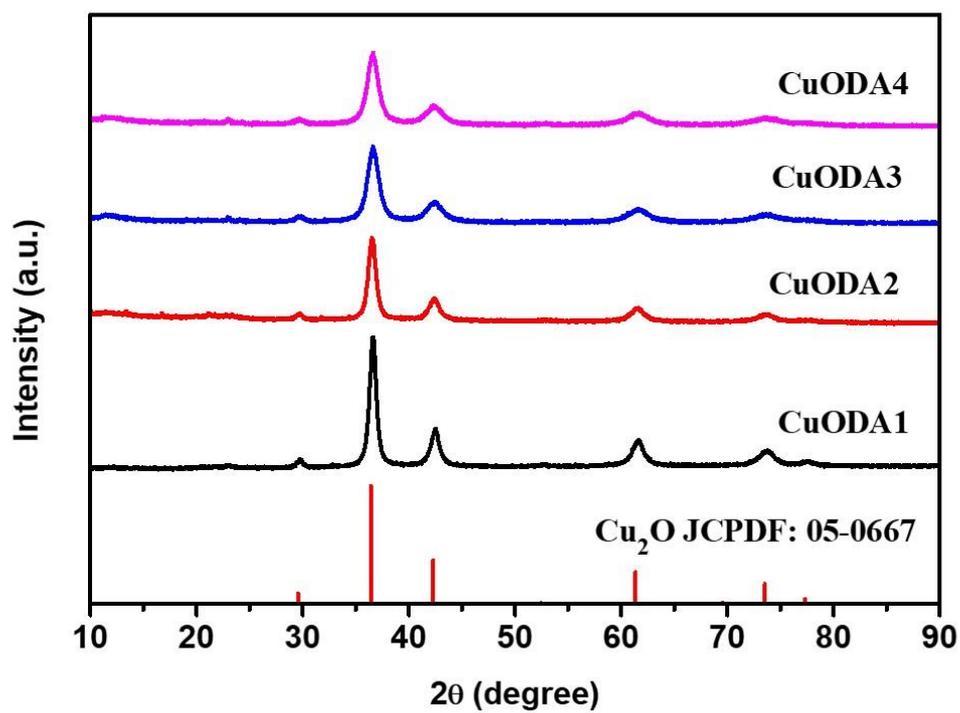

**Figure S1.** PXRD patterns of CuODA1, CuODA2, CuODA3 and CuODA4. The standard pattern of Cu$_2$O (PDF #05-0667) is also included.

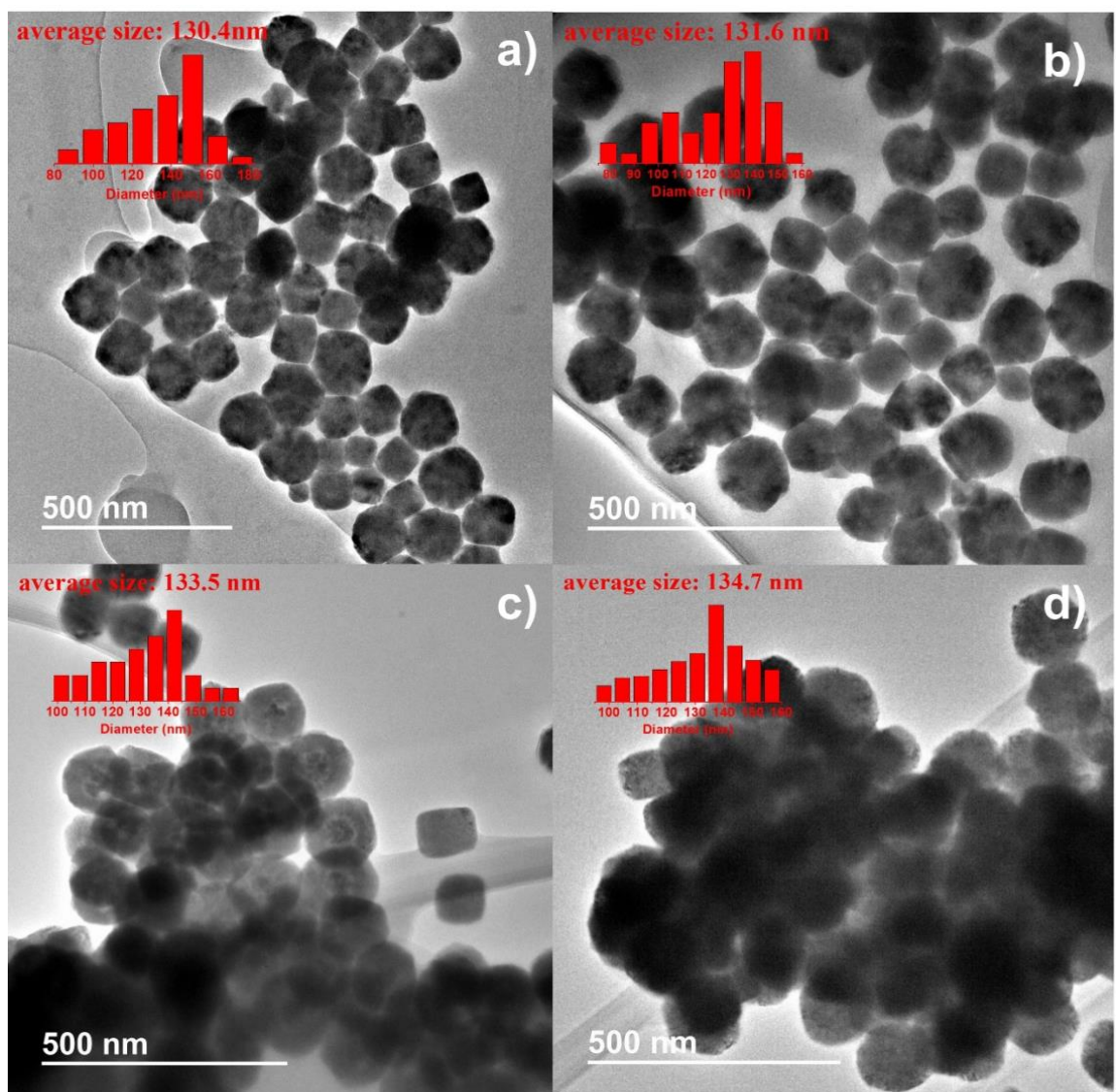

**Figure S2**. TEM images of CuODA1 (a), CuODA2 (b), CuODA3 (c) and CuODA4. Insets show histogram of particle size distribution and the average particle size.

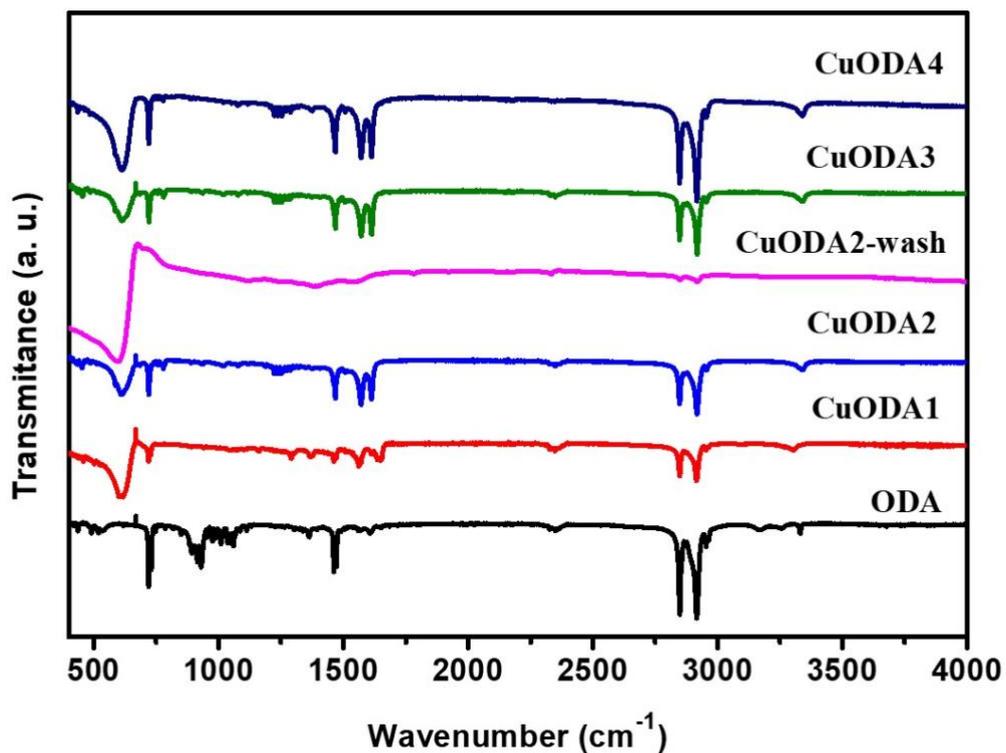

**Figure S3**. ATR-FTIR spectra of ODA, CuRN1, CuRN2, CuRN3 and CuRN4.

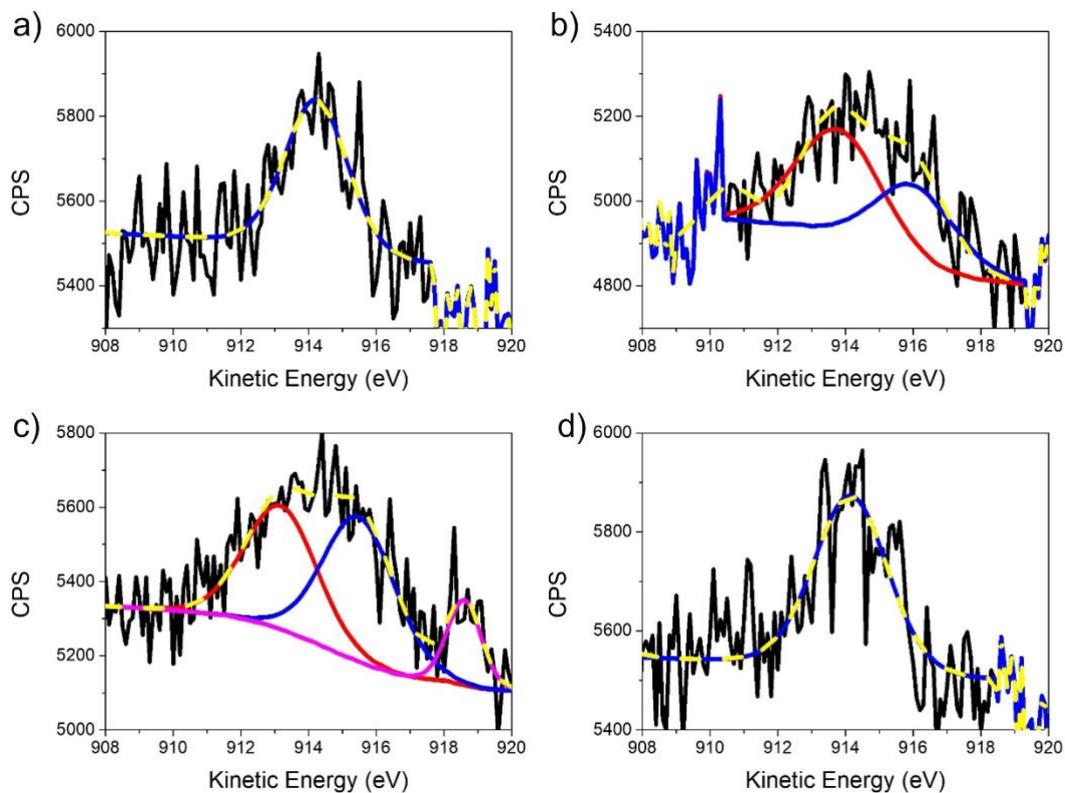

**Figure S4**. Cu LMM Auger spectra of CuODA1 (a), CuODA2 (b), CuODA3 (c) and CuODA4 (d).

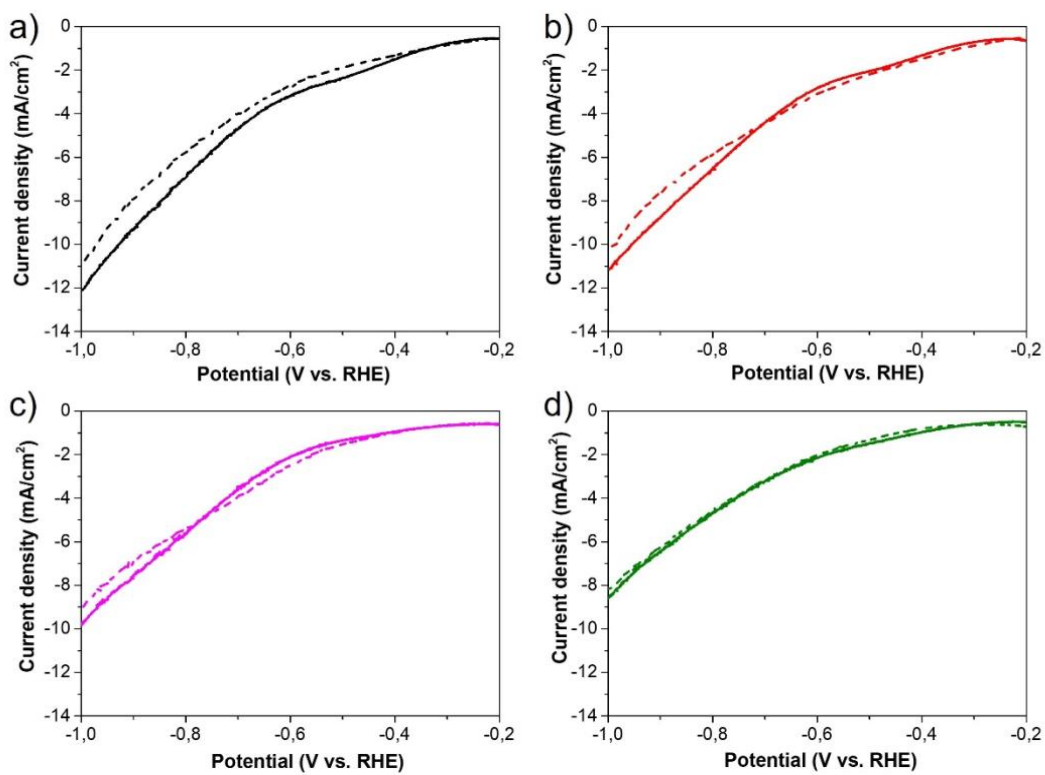

**Figure S5**. Liner Sweep Voltammetry (LSV) of CuODA1 (a), CuODA2 (b), CuODA3 (C) and CuODA4 (d) in $N_2$-saturated (dashed) and $CO_2$-saturated (solid) 1 M $KHCO_3$ electrolyte.

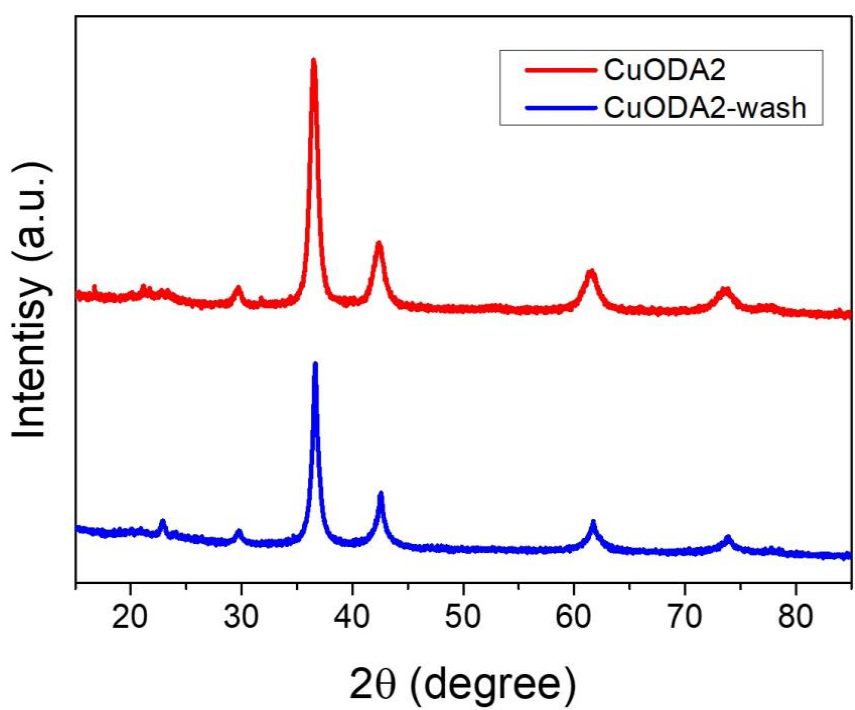

**Figure S6**. PXRD patterns of CuODA2 and CuODA2-wash.

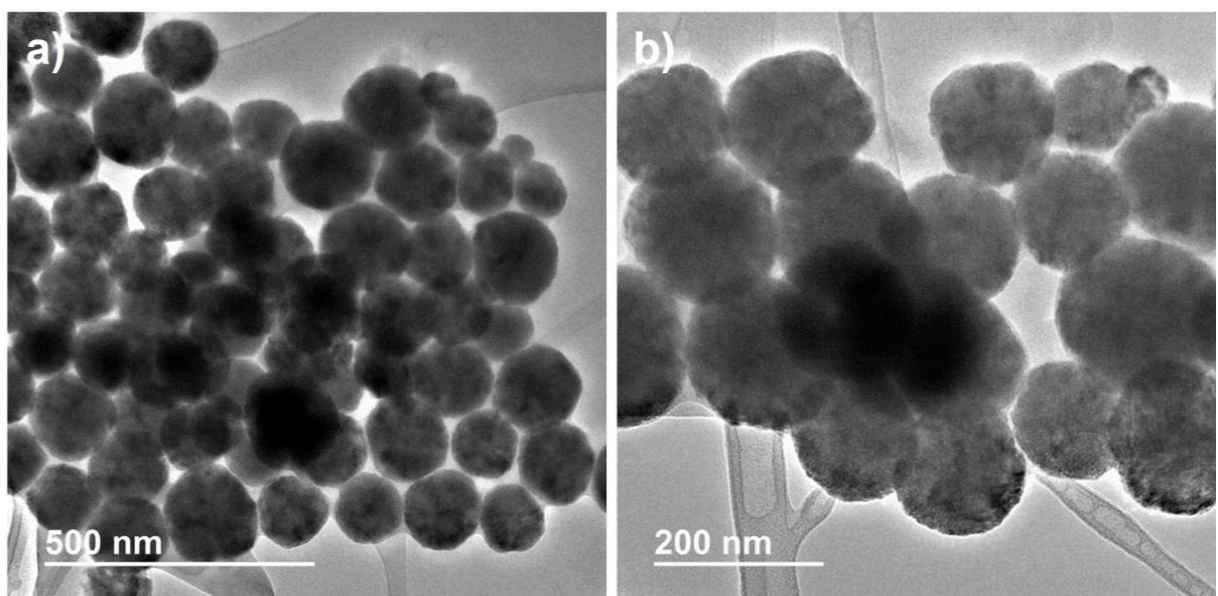

**Figure S7**. TEM images of CuODA2-wash samples acquired at different magnifications.

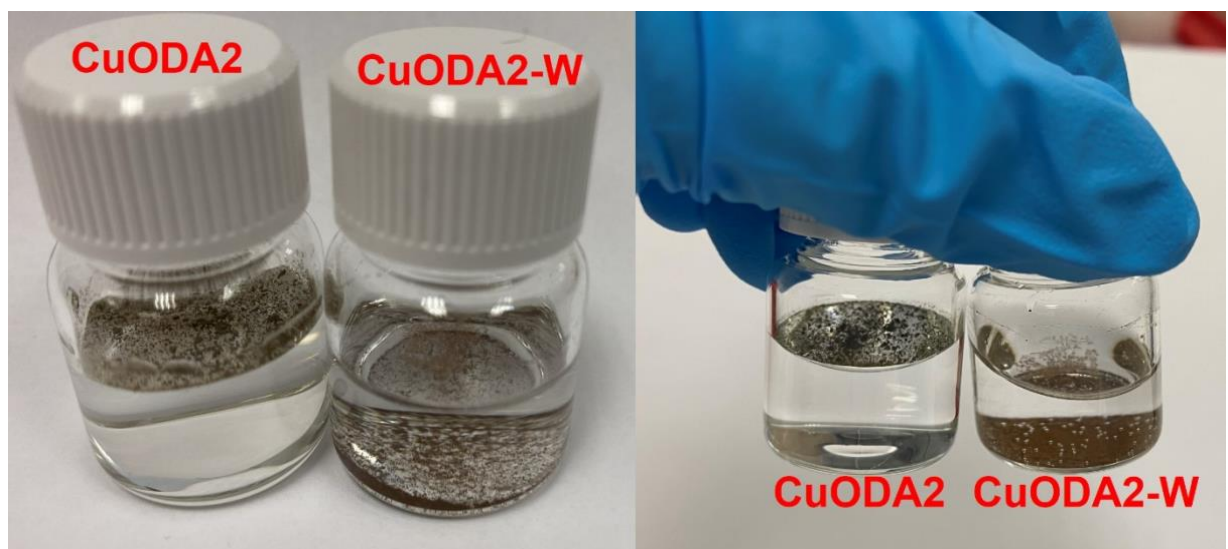

*Figure S8.* *H₂O hydrophobicity test of CuODA2 and CuODA2-wash. It can be seen that CuODA2-wash precipitates at the bottom, while CuODA2 floats on water after 2 h.*

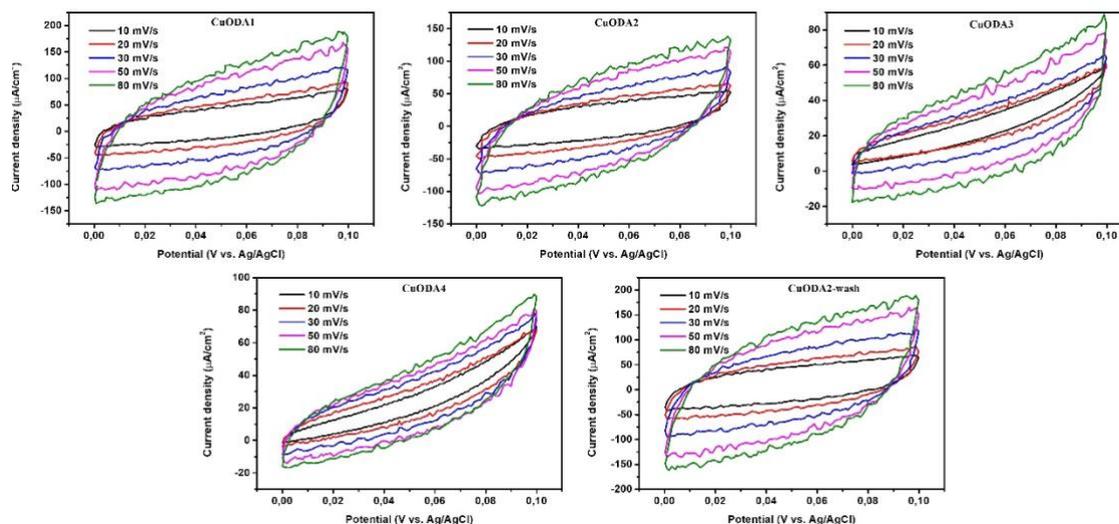

**Figure S9.** Cyclic voltammograms of the samples at different scan rates from 10 to 80 mV/s.

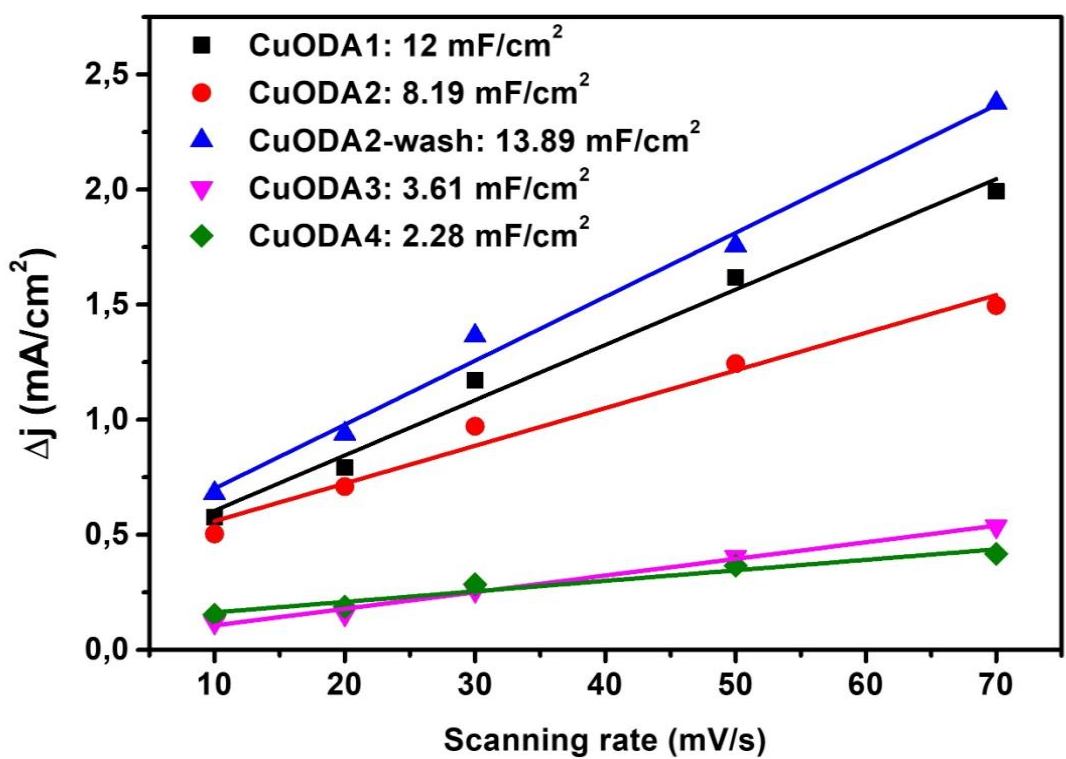

**Figure S10.** Capacitive current as scan rate function of the different samples. The $C_{DL}$ obtained is indicated for each sample.

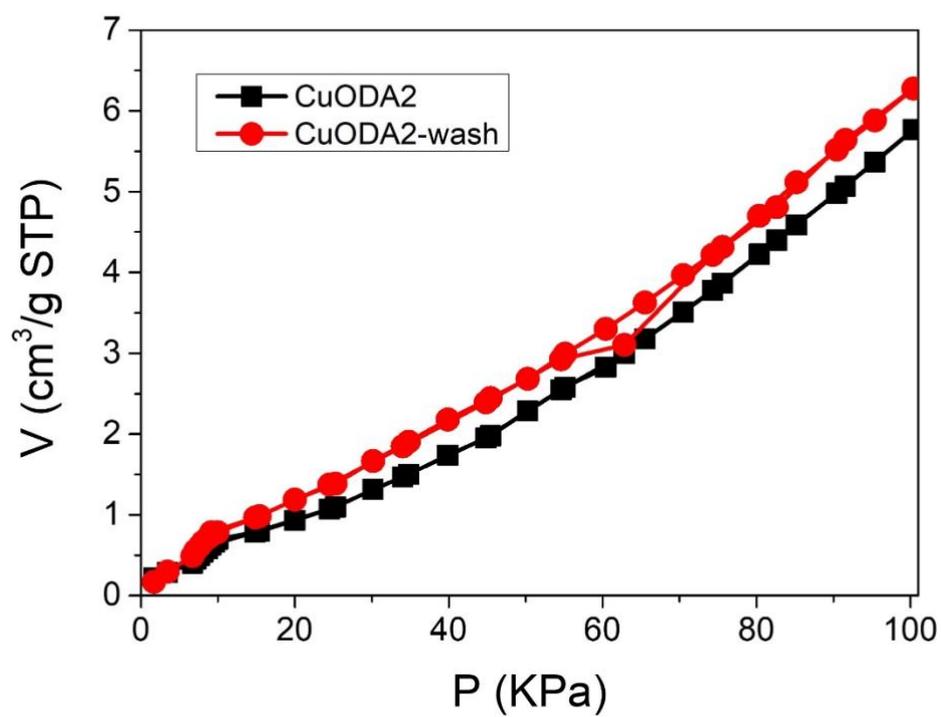

*Figure S11*. $CO_2$ *adsorption isotherms at 25 ºC of CuODA2 (black) and CuODA2-wash (red).*

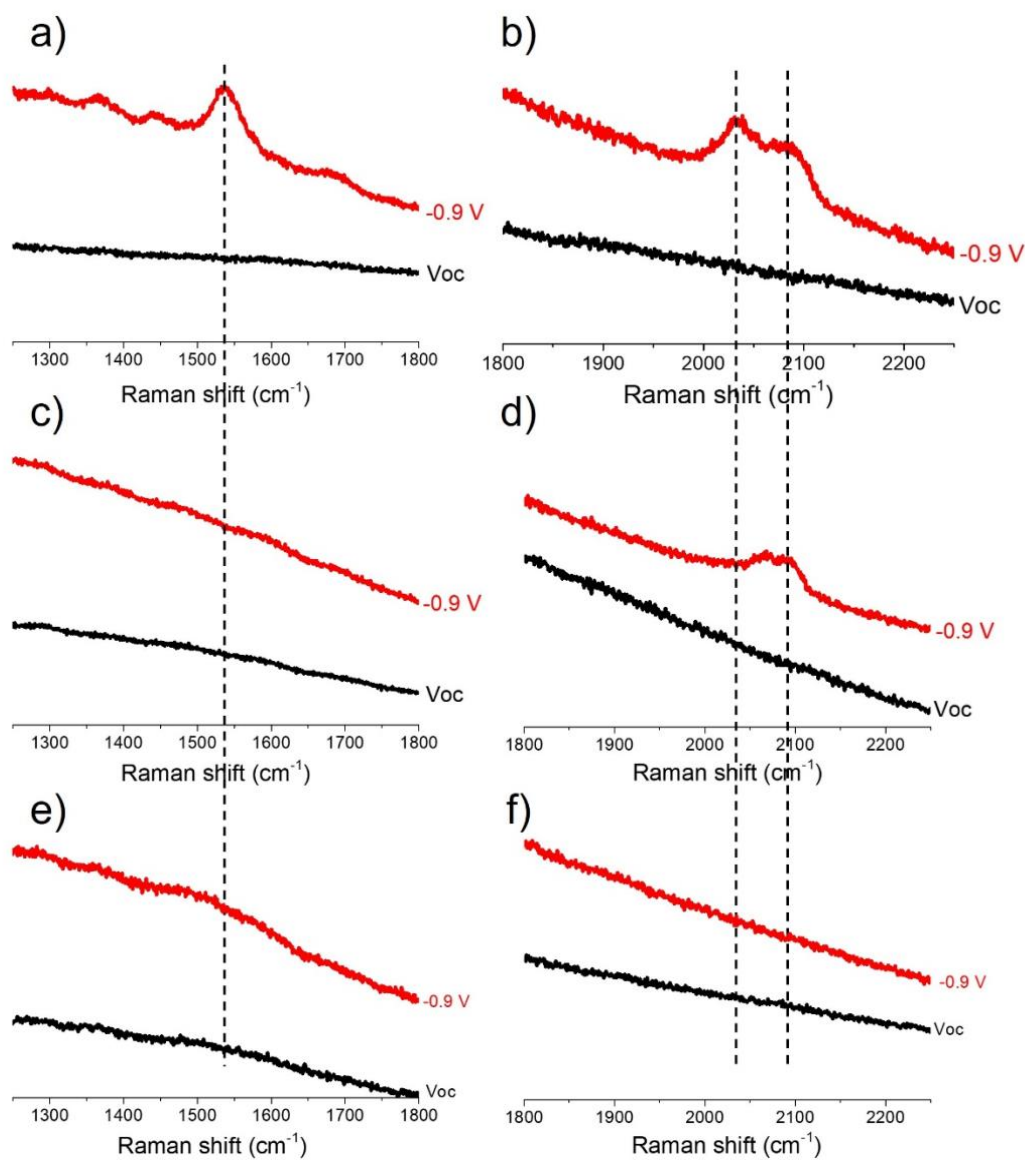

**Figure S12.** In situ electrochemical Raman spectra of CuODA2 (a and b), CuODA4 (c and d) and CuODA2-wash (e and f) obtained at open circuit potential (black) and at cathode potential of -0.9 V (red). Laser excitation 785 nm.

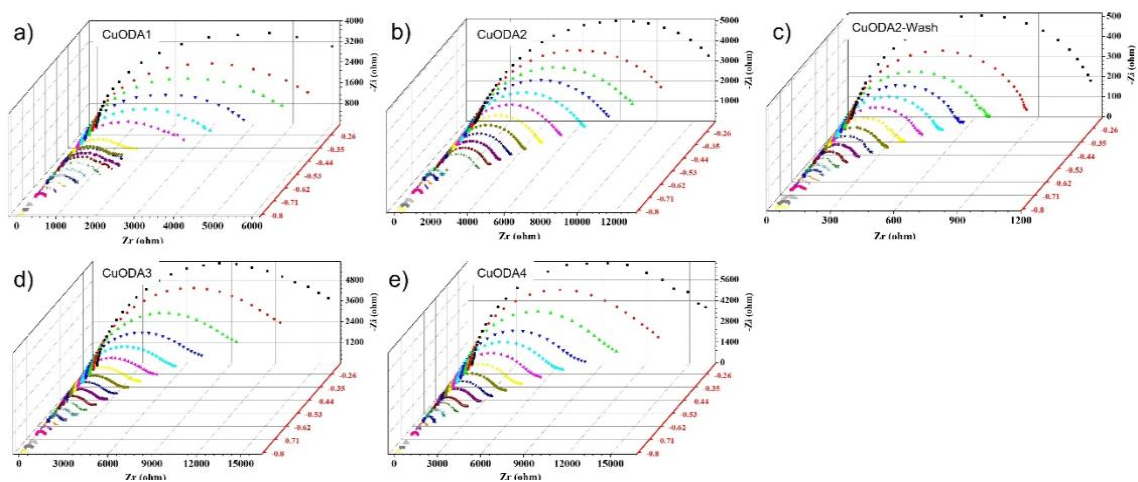

**Figure S13**. Nyquist plots of CuODA1 (a), CuODA2 (b), CuODA2-wash (c), CuODA3 (d) and CuODA4 (e) collected at 20 different potentials between -0.2 to -0.8 V vs. RHE in 0.03 V increments. The spectra were collected from 0.5 Hz to 30 kHz at 10 points per decade. $CO_2$-saturated 1M $KHCO_3$ electrolyte. Pt wide and KCl-saturated Ag/AgCl were used as counter and reference electrodes, respectively.

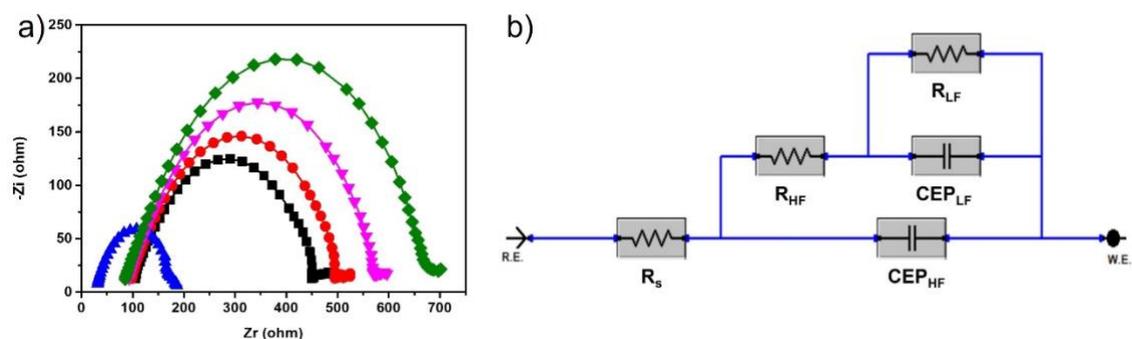

**Figure S14**. a) Nyquist plot of CuODA2 measured at -0.7 V vs. RHE in the range from 0.5 Hz to 30 kHz. b) Equivalent circuit used to fit the obtained experimental data.

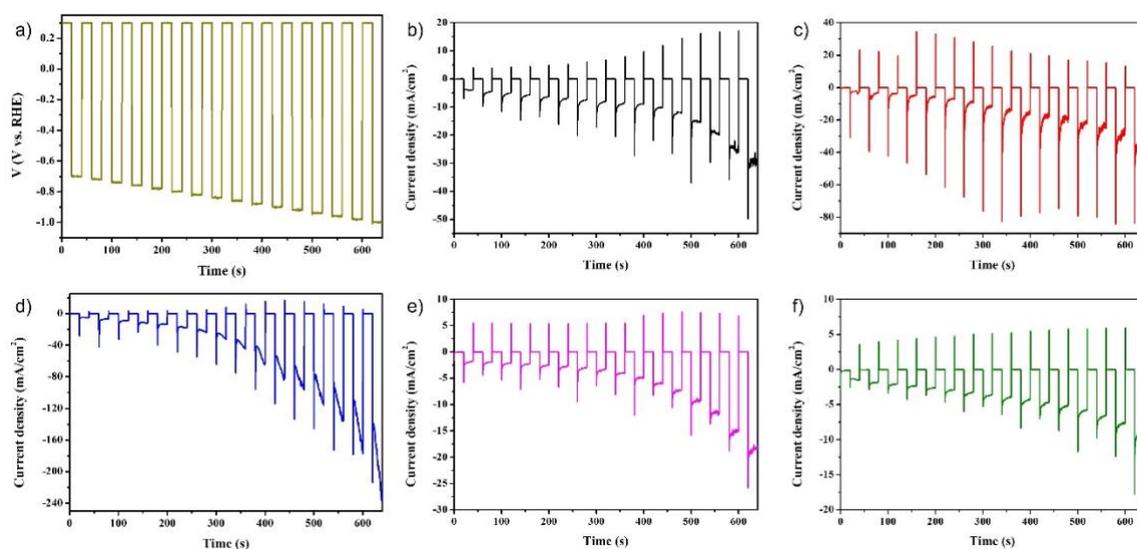

**Figure S15**. (a) Stepwise potential profile applied to the working electrode in pulsed voltammetry experiments. (b-f) Pulse responses for CuODA1 (b), CuODA2 (c), CuODA3 (d), CuODA4 (e) and CuODA2-wash (f).

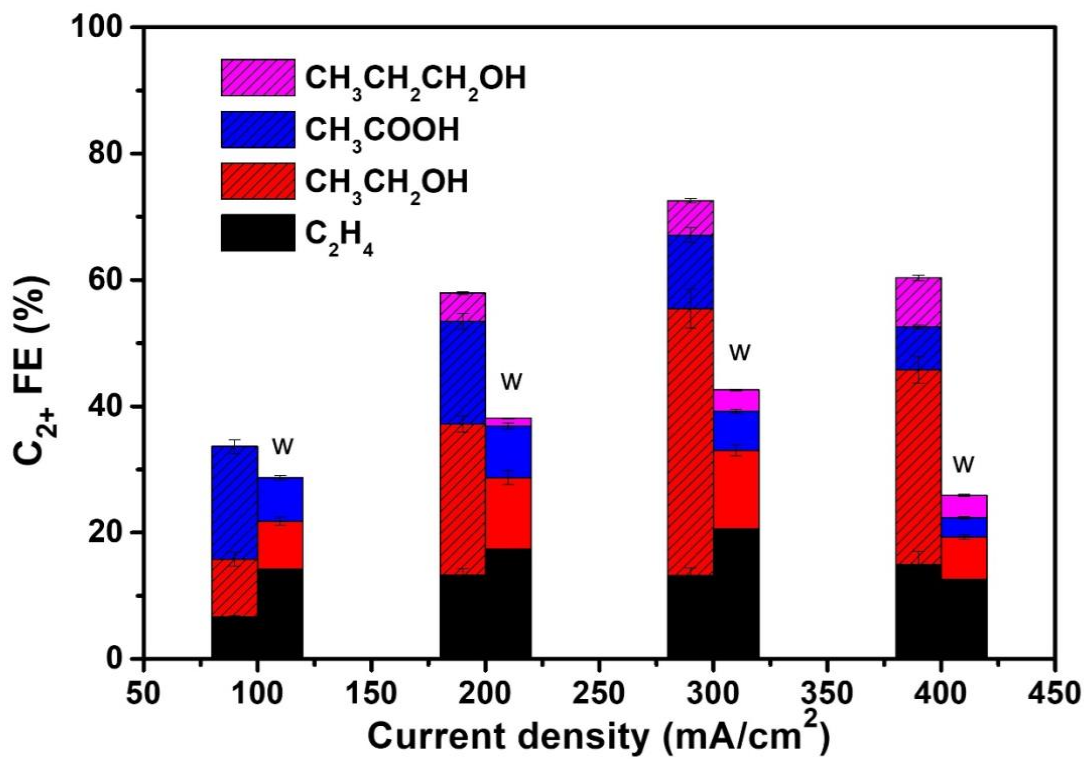

**Figure 16**. FE for $CH_2=CH_2$ (black), $CH_3CH_2OH$ (red), $CH_3COOH$ (blue) and $CH_3CH_2CH_2OH$ (pink) obtained from CuODA2 (solid bars) and CuODA2-wash (branded bars; w) at different current densities and using a flow electrochemical cell.

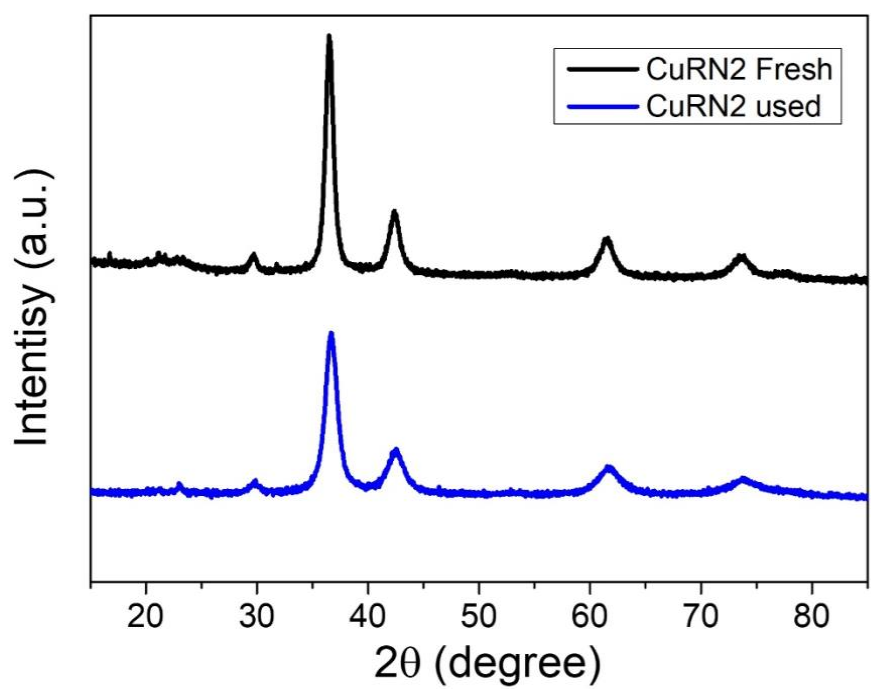

**Figure 17**. XRD patterns of CuODA2 before (black) and after (blue) continuous flow operation at 300 mA/cm² for 2 h.

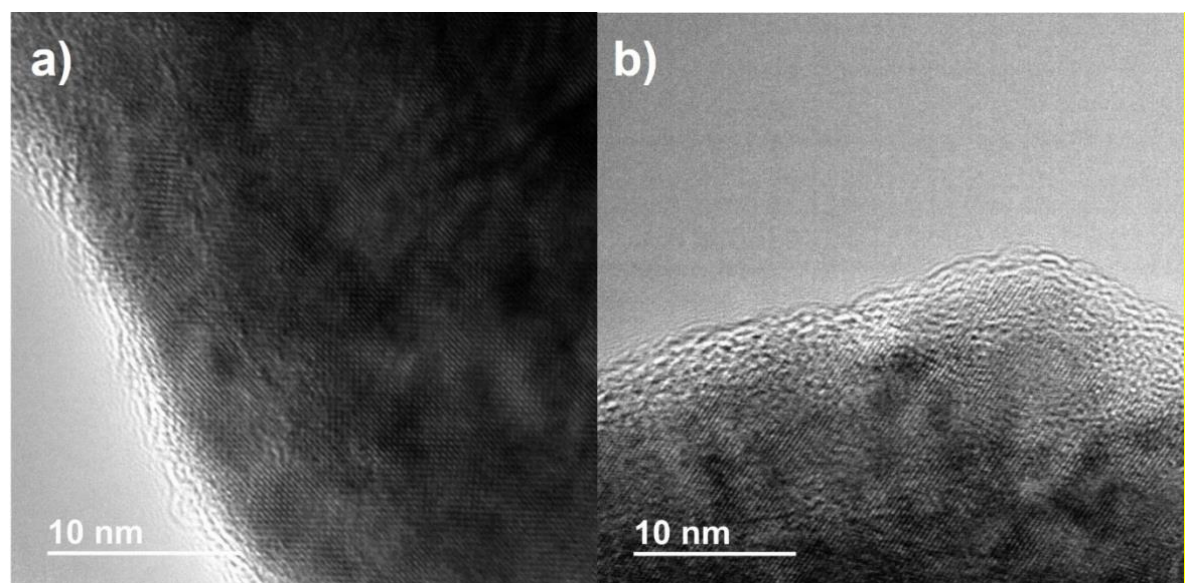

**Figure S18**. HRTEM images of CuODA2 after reaction.

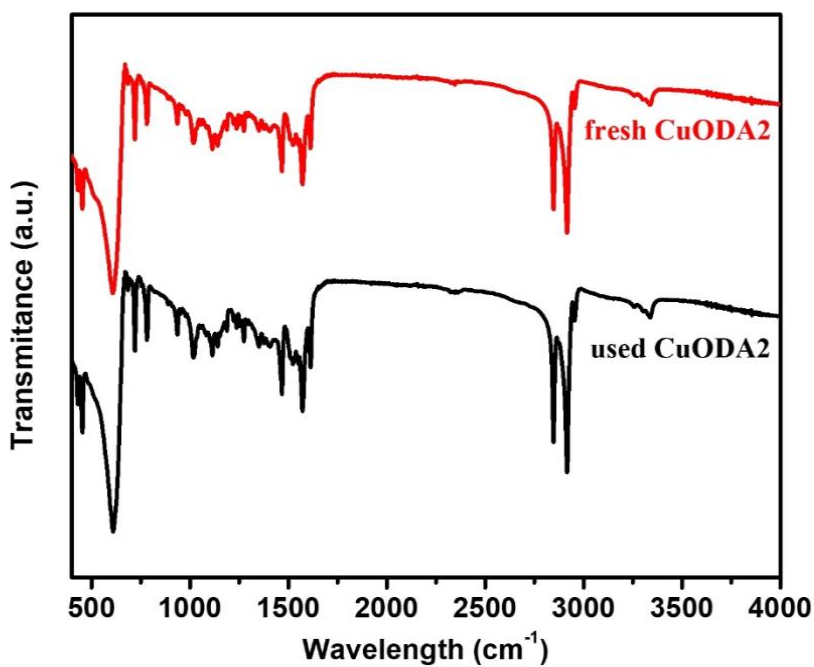

**Figure S19**. FTIR spectrum of fresh (red) and after reaction (black) CuODA2 catalyst.

## References

[1]  D. Wakerley, S. Lamaison, F. Ozanam, N. Menguy, D. Mercier, P. Marcus, M. Fontecave, V. Mougel, *Nature Materials* **2019**, 18, 1222.
[2]  Y. Zhang, X.-Y. Zhang, K. Chen, W.-Y. Sun, *ChemSusChem* **2021**, 14, 1847.
[3]  X. Chen, J. Chen, N. M. Alghoraibi, D. A. Henckel, R. Zhang, U. O. Nwabara, K. E. Madsen, P. J. A. Kenis, S. C. Zimmerman, A. A. Gewirth, *Nature Catalysis* **2021**, 4, 20.
[4]  C. Y. J. Lim, M. Yilmaz, J. M. Arce-Ramos, A. D. Handoko, W. J. Teh, Y. Zheng, Z. H. J. Khoo, M. Lin, M. Isaacs, T. L. D. Tam, Y. Bai, C. K. Ng, B. S. Yeo, G. Sankar, I. P. Parkin, K. Hippalgaonkar, M. B. Sullivan, J. Zhang, Y.-F. Lim, *Nature Communications* **2023**, 14, 335.

[5] F. Li, A. Thevenon, A. Rosas-Hernández, Z. Wang, Y. Li, C. M. Gabardo, A. Ozden, C. T. Dinh, J. Li, Y. Wang, J. P. Edwards, Y. Xu, C. McCallum, L. Tao, Z.-Q. Liang, M. Luo, X. Wang, H. Li, C. P. O'Brien, C.-S. Tan, D.-H. Nam, R. Quintero-Bermudez, T.-T. Zhuang, Y. C. Li, Z. Han, R. D. Britt, D. Sinton, T. Agapie, J. C. Peters, E. H. Sargent, *Nature* **2020**, 577, 509.

[6] Z. Han, R. Kortlever, H.-Y. Chen, J. C. Peters, T. Agapie, *ACS Central Science* **2017**, 3, 853.

[7] A. Ozden, F. Li, F. P. García de Arquer, A. Rosas-Hernández, A. Thevenon, Y. Wang, S.-F. Hung, X. Wang, B. Chen, J. Li, J. Wicks, M. Luo, Z. Wang, T. Agapie, J. C. Peters, E. H. Sargent, D. Sinton, *ACS Energy Letters* **2020**, 5, 2811.

[8] F. Li, Y. C. Li, Z. Wang, J. Li, D.-H. Nam, Y. Lum, M. Luo, X. Wang, A. Ozden, S.-F. Hung, B. Chen, Y. Wang, J. Wicks, Y. Xu, Y. Li, C. M. Gabardo, C.-T. Dinh, Y. Wang, T.-T. Zhuang, D. Sinton, E. H. Sargent, *Nature Catalysis* **2020**, 3, 75.

[9] L. Yang, X. Lv, C. Peng, S. Kong, F. Huang, Y. Tang, L. Zhang, G. Zheng, *ACS Central Science* **2023**, 9, 1905.

[10] S. Zhao, O. Christensen, Z. Sun, H. Liang, A. Bagger, K. Torbensen, P. Nazari, J. V. Lauritsen, S. U. Pedersen, J. Rossmeisl, K. Daasbjerg, *Nature Communications* **2023**, 14, 844.

[11] H. Wu, J. Li, K. Qi, Y. Zhang, E. Petit, W. Wang, V. Flaud, N. Onofrio, B. Rebiere, L. Huang, C. Salameh, L. Lajaunie, P. Miele, D. Voiry, *Nature Communications* **2021**, 12, 7210.

[12] S. Chen, C. Ye, Z. Wang, P. Li, W. Jiang, Z. Zhuang, J. Zhu, X. Zheng, S. Zaman, H. Ou, L. Lv, L. Tan, Y. Su, J. Ouyang, D. Wang, **2023**, 62, e202315621.